\documentclass{article}

\def\myputfigure#1#2#3#4%
{\hskip0.03\textwidth%
\hfill\makebox[0pt]{\hskip#2in\includegraphics[width=#3\textwidth]{#1}}\vskip#4pt\hfill}

\usepackage{graphicx,emulateapj}

\def\MICM{\hbox{$M_{ICM}$}}
\def\fICM{\hbox{$f_{ICM}$}}
\def\TX{\hbox{$\left<T_X\right>$}}

\def\rave{\hbox{$\left<r\right>/r_{500}$}}
\def\rfive{\hbox{$r_{500}$}}
\def\mfive{\hbox{$M_{500}$}}
\def\betaeff{\hbox{$\beta_{eff}$}}

\def\spose#1{\hbox to 0pt{#1\hss}}
\def\lta{\mathrel{\spose{\lower 3pt\hbox{$\mathchar"218$}}
     \raise 2.0pt\hbox{$\mathchar"13C$}}}
\def\gta{\mathrel{\spose{\lower 3pt\hbox{$\mathchar"218$}}
     \raise 2.0pt\hbox{$\mathchar"13E$}}}
\def\tCDM{$\tau$CDM}

\begin{document}

\submitted{To appear in The Astrophysical Journal: submitted Oct 2, '98; 
accepted Dec 15, '98}

\title{Properties of the Intracluster Medium in an Ensemble of
Nearby Galaxy Clusters}

\author{Joseph J. Mohr\altaffilmark{1,2,3,4} Benjamin Mathiesen\altaffilmark{2}
\& August E. Evrard\altaffilmark{2}}
 
\affil{
$^1$Department of Astronomy and Astrophysics, University of Chicago, Chicago, IL  60637 \\
$^2$Department of Physics \& $^3$Department of Astronomy, University of Michigan, Ann Arbor, MI 48109}
\altaffiltext{4} {AXAF Fellow}
 
\authoremail{mohr@oddjob.uchicago.edu}
\authoremail{bfm@umich.edu}
\authoremail{evrard@umich.edu}
 
\begin{abstract}
We present a systematic analysis of the intracluster medium (ICM)
in an X-ray flux limited sample of 45 galaxy clusters.
Using archival ROSAT PSPC data and published ICM temperatures, we present
best fit double and single $\beta$ model profiles, and extract ICM 
central densities and radial distributions.  We use the data and an 
ensemble of numerical cluster simulations to quantify 
sources of uncertainty for all reported parameters.

We examine the ensemble properties within the context of
models of structure formation and feedback from galactic winds.
We present best fit ICM mass-temperature \MICM-\TX\ 
relations for \MICM\ calculated within \rfive\ and 1$h^{-1}_{50}$~Mpc.
These relations exhibit small scatter (17\%), 
providing evidence of regularity
in large, X-ray flux limited cluster ensembles.  Interestingly, the
slope of the \MICM-\TX\ relation (at limiting radius \rfive) is steeper
than the self-similar expectation by 4.3$\sigma$.
We show that there is a mild dependence of ICM mass fraction \fICM\ on \TX;
the clusters with ICM temperatures below 5~keV have
a mean ICM mass fraction $\left<f_{ICM}\right>=0.160\pm0.008$ which is
significantly lower than that of the hotter clusters 
$\left<f_{ICM}\right>=0.212\pm0.006$ (90\% confidence intervals).  
In apparent contradiction with previously published analyses, our large,
X-ray flux limited cluster sample provides
no evidence for a more extended radial ICM distribution in low \TX\ 
clusters down to the sample limit of 2.4\,keV.  

By analysing simulated clusters we find that density variations enhance the
cluster X-ray emission and cause \MICM\ and \fICM\ to be overestimated by
$\sim$12\%.  Additionally, we use the simulations to estimate an \fICM\ 
depletion factor at \rfive. We use the bias corrected  mean \fICM\ within
the hotter cluster subsample as a lower limit on the cluster baryon fraction.
In combination with nucleosynthesis constraints this measure provides a firm
upper limit on the cosmological density parameter for clustered matter
$\Omega_M\le(0.36\pm0.01)h^{-1/2}_{50}$.
\end{abstract}


\section{INTRODUCTION}
The properties of galaxy cluster virial regions provide powerful
constraints on models of structure formation and evolution.  
For example, evidence for continuing accretion in nearby clusters 
constrains the cosmological density parameter $\Omega_0$ and the 
slope of the power spectrum of density perturbations $P(k)$ on cluster scales.
Combining the baryon fraction within cluster virial regions 
with nucleosynthesis constraints on the baryon to photon ratio
provides an estimate of the cosmological density parameter 
for clustered matter $\Omega_M$.
An emerging theoretical consensus that dissipationless collapse produces haloes
which conform to a ``universal'' density profile (\cite{navarro97}) implies
that measurements of cluster virial structure
provide constraints on the nature of dark matter.
Therefore, systematic, high precision analyses of large, well
defined cluster samples can potentially enjoy broad and lasting impact.  

There are several tools available to probe galaxy cluster virial regions. 
One can use gravitational lensing to study the 
characteristics of the clustered mass directly.  
At the present time, typical cluster weak lensing maps
have peak signal to noise of $\sim$10 even
at disappointingly low angular
resolutions (e.g. Fischer \& Mohr in prep),
making it difficult to resolve differences
between observed profiles and the NFW prediction.
Studies of strong lensing can provide
extremely detailed constraints on the structure of cluster cores 
(\cite{tyson98}), but the required alignments of lens, source and
observer occur in few clusters.  Dynamical studies using galaxies 
can potentially resolve
the dark matter structure and galactic orbits simultaneously, but doing
so requires such a large galaxy sample that data from multiple clusters
must be rescaled and stacked (e.g. \cite{carlberg97}), precluding 
tests of anything except average properties of cluster mass profiles.
Archival ROSAT PSPC X-ray images of nearby galaxy clusters are of
sufficient quality that high precision studies of the ICM in individual
clusters are possible.  Moreover, it is straightforward to study a large, 
well defined cluster ensemble, making it possible to compare the 
structure of high and low mass clusters and to {\it quantify} departures 
from self similarity. 

Here we present a precision analysis of the intracluster medium in
45 nearby clusters.
Our goals include (i) a study of cluster regularity,
(ii) a quantitative characterization of the ICM mass-temperature (\MICM-\TX)
relation for comparison to future structure formation and galaxy
formation simulations, (iii) a definitive test of whether
the ICM mass fraction \fICM\ and radial distribution vary with cluster mass,
and (iv) an improved measure of $\left<f_{ICM}\right>$, leading to an
upper limit on the cosmological density parameter $\Omega_M$.

Our technique is a synthesis of profile fitting methods
applied to elliptical galaxies (\cite{saglia93,mohr97b})
and procedures applied to examine the ICM in smaller cluster
samples (\cite{david93,david95}).
The primary observational data required to study each cluster are
(i) the X-ray surface brightness profile, which constrains the 
radial distribution of the ICM along with (ii) the X-ray
luminosity within some aperture
and (iii) the emission weighted mean ICM temperature \TX,
which together constrain the ICM central density.  We demonstrate that
our ICM constraints are insensitive to temperature variations 
within the cluster.
Departing from some other published analyses, we do not assume
a particular form for the dark matter distribution, 
because it is not necessary to do so when studying the ICM; however,
we do assume a spherically symmetric ICM distribution. We test our technique
on an ensemble of hydrodynamical cluster simulations and quantify
the effects of present epoch mergers and asphericity on our results.

We describe the data and numerical simulations in $\S2$, detail the analysis in
$\S3$, and then present the results in $\S4$.  Section 5 contains
a discussion of the results.  We use $H_0=50$~km/s/Mpc and $q_0={1\over2}$
throughout.

\section{DATA}

We study members of an X-ray flux limited sample of 55 clusters (\cite{edge90})
that were observed with the ROSAT PSPC.
Archival PSPC images of forty-eight members exist, but Abell\,399
lies partially outside the PSPC field of view (FOV),
Virgo extends well beyond the PSPC FOV, and the quality of the short
(850~s) exposure of Abell\,2147 image is well below the others. 
Of the remaining 45 clusters, three have no upper limits on
their temperature uncertainties, and so as noted later we exclude them
from portions of the final analysis. The final sample
is very similar to the same sample analyzed in a study of the
X-ray Size-Temperature Relation (\cite{mohr97a}).

\subsection{X-ray Images}

We obtain the ROSAT data products on-line through the High Energy Astrophysics
Science Archive Research Center (HEASARC), and we reduce these images using
PROS and Snowden analysis software (\cite{snowden94}).  The X-ray image
for each PSPC observation is the sum of individually flat fielded
Snowden energy bands $R4$ through $R7$, corresponding approximately to 
the energy band 0.5--2.0\,keV.  We exclude time intervals with master veto
rates higher than 220\,cts/s, and exclude other high background time
intervals whose inclusion would degrade the detection significance of a source
10\% as bright as the background (\cite{pildis95}). We produce uncertainty
images by assuming Poisson noise, estimating a 1\% uncertainty in the
flatfields, and applying standard error propagation.  Multiple PSPC
observations of the same cluster are registered and combined using the
positions of bright point sources where possible, or, alternatively the
image header pointing positions.  We remove obvious point sources not
coincident with the peak in the extended emission using the IRAF
task {\it imedit}.  The pixel scale is 14\farcs947, and the effective
angular resolution is FWHM$\sim$0\farcm75.  Table \ref{basicdata}
contains a list of clusters and relevant information, including
ROSAT sequence number(s) of observations incorporated into the
final cluster image, effective exposure time $t_{exp}$, and the emission
weighted mean ICM temperature \TX.

The procedure described above produces images in units of photons per second;
we use PROS software to convert to cgs flux units ergs/s/cm$^2$ in the
restframe of the cluster.  In the
conversion we assume an isothermal ICM emitting a Raymond--Smith
spectrum with 30\% solar abundances and a temperature \TX.
We also correct for galactic absorption.

\myputfigure{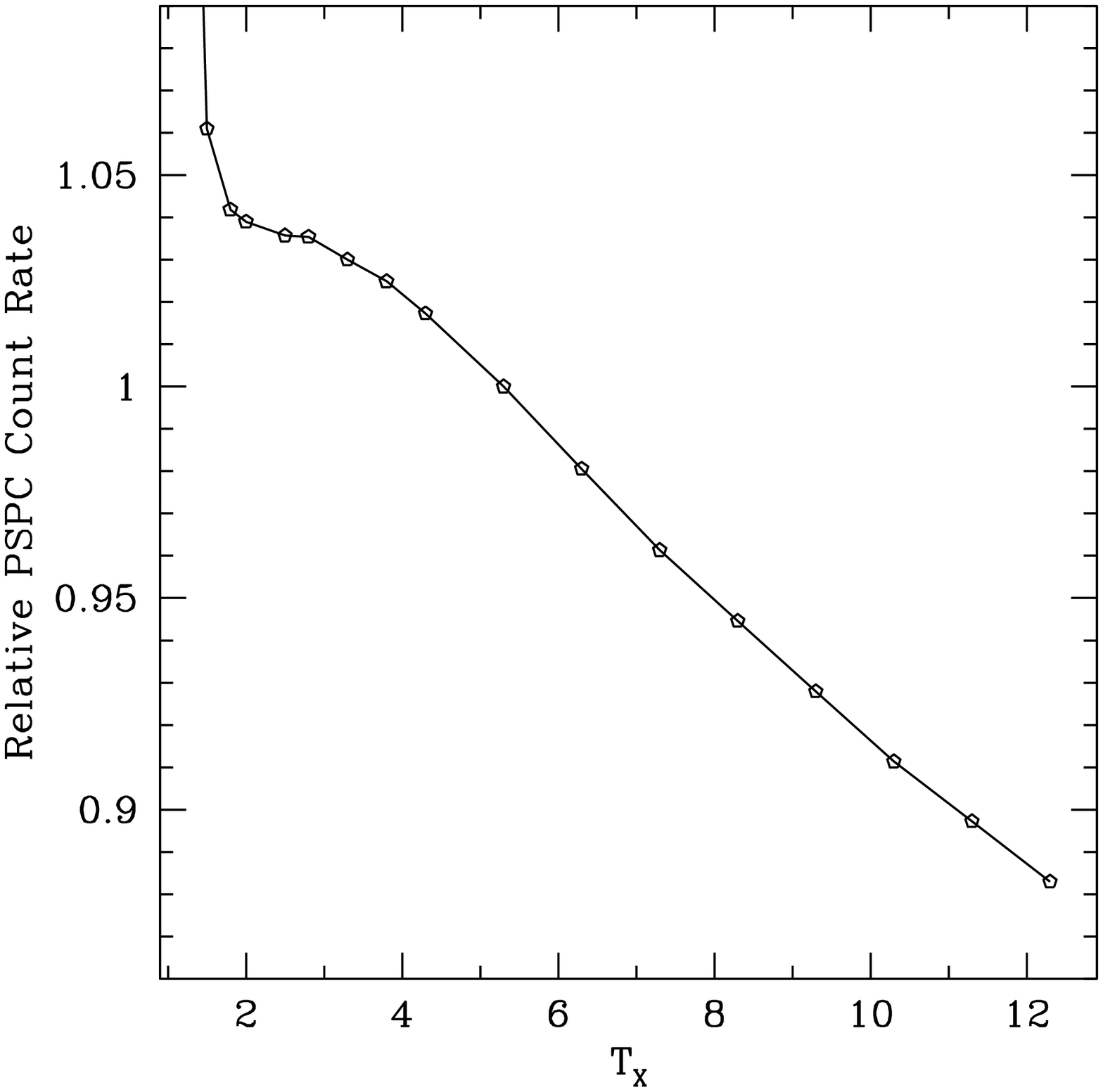}{-3.5}{0.50}{-30}\figcaption{
We plot the relative PSPC count rate in the 0.5-2.0\,keV
band for a parcel of gas with constant emission measure and
varying temperature.  The emission model is Raymond-Smith with
30\% solar abundances.  Note that for $T_X>1.5$\,keV a factor
of two increase in temperature only decreases the PSPC
count rate by $\sim$10\%.}
\label{PSPCemis}
\medskip

\subsection{ICM Temperatures}

We use published, emission weighted ICM temperatures \TX\ in our analysis
(see Table \ref{basicdata}).
Approximately a third of these temperatures come from {\it Einstein} MPC
observations, but where possible we have substituted
more accurate measurements made with {\it Ginga} or ASCA observations. 
The sources for the temperature measurements include:
{\it Einstein} MPC: \cite{david93};
{\it Ginga}: \cite{day91,allen92,johnstone92,david93,hughes93,arnaud98};
ASCA: \cite{henrik96,marke96,matsu96,tamura96,marke97,marke98}.

\subsection{Hydrodynamical Cluster Simulations}

We use an ensemble of 48 hydrodynamical cluster simulations to better
understand the systematics in the measurements presented below.
The cluster simulations are carried out within four different cold dark matter 
(CDM) dominated cosmogonies
(1) SCDM  ($\Omega=1$,     $\sigma_8=0.6$, $h_{50}=1$, $\Gamma=0.5$),
(2) \tCDM ($\Omega=1$,     $\sigma_8=0.6$, $h_{50}=1$, $\Gamma=0.24$),
(3) OCDM  ($\Omega_0=0.3$, $\sigma_8=1.0$, $h_{50}=1.6$, $\Gamma=0.24$), and
(4) LCDM  ($\Omega_0=0.3$, $\lambda_0=0.7$, $\sigma_8=1.0$, $h_{50}=1.6$, $\Gamma=0.24$).
Here $\sigma_8$ is the power spectrum normalization on $8 h^{-1}$~Mpc scales;
initial conditions are Gaussian random fields consistent with a 
CDM transfer function with the specified $\Gamma$ (\cite{davis85}).
Within each of these models, we use two $128^3$
N--body only simulations of cubic regions with scale $\sim400$~Mpc to
determine sites of cluster formation.  Within these initial runs the
virial regions of clusters with Coma--like masses of $10^{15}$~M$_\odot$
contain $\sim$10$^3$ particles.

Using the N--body results for each model,
we choose clusters for additional study.
We zoom in on these clusters, resimulating them at higher resolution with
gas dynamics and gravity on a $64^3$ N--body grid.  The large wavelength
modes of the initial density field are sampled from the initial conditions of
the large scale N--body simulations, and power on smaller scales is sampled
from the appropriate CDM power spectrum.  The simulation scheme is
P3MSPH (\cite{evrard88}), the baryon density is a fixed fraction of the
total $\Omega_b = 0.2 \Omega_0$, and radiative cooling is ignored. 

Simulating individual clusters requires two steps:  (1) an initial, $32^3$,
purely N--body simulation to identify which portions of the initial density
field lie within the cluster virial region at the present epoch, and (2) a
final, effectively $64^3$, three species, hydrodynamical simulation. 
In the final simulation the portion of the initial density field which ends up
within the cluster virial region by the present epoch is represented using
dark matter and gas particles of equal number, while the
portions of the initial density field that do not end up within the cluster
virial region by the present epoch are represented using a third,
collisionless, high mass, species.  The high mass species is
8 times more massive than the dark matter particles
in the central, high resolution region.
This approach allows us to include the tidal effects of the surrounding
large scale structure and the gas dynamics of the cluster virial region with
simulations that take only a few days of CPU time on a low end UltraSparc.
The scale of the 
simulated region surrounding each cluster is in the range 50--100~Mpc, and
varies as $M_{halo}^{1/3}$, where $M_{halo}$ is approximately the mass enclosed
within the present epoch turn around radius.  Thus, the 48 simulated clusters
in our final sample have similar fractional mass resolution; the spatial
resolution varies from 125--250~kpc. 
The masses of the final cluster sample vary by an order of magnitude.
We create X--ray images and temperature
maps for further analysis following procedures described in Evrard (1990).
 
\section{ANALYSIS}
In this section, we detail our data reduction procedures,
placing special emphasis on the determination of statistical uncertainties and
systematic errors in the derived cluster ICM gas masses.

\subsection{Modeling the X-ray Surface Brightness Profile}

We use the reduced, cluster X--ray images to constrain the underlying ICM
distribution.  Our approach is to fit radial surface brightness 
profiles of the cluster emission
to the standard $\beta$ model (\cite{cava78}) of the form
\begin{equation}
I(R) = I_0 \left[1 + {\left(R\over R_c\right)}^2\right]^{-3\beta+{1\over2}}
\label{beta}
\end{equation}
where $R_c$ is the core radius and $r^{-3\beta}$ is the
asymptotic radial fall-off of the underlying ICM density distribution.
Because the PSPC photon detection rate for a parcel of gas with
constant emission measure varies only moderately with temperature above
1.5\,keV (see Fig. \ref{PSPCemis}), the distribution of X--ray
photons in PSPC images directly
constrains the ICM density distribution regardless of the temperature
structure (see similar discussion of the {\it Einstein} IPC: \cite{fabric80}).
Fitting the $\beta$ model provides a systematic approach to studying the
ICM properties.
In the case where cluster X--ray emission is well described by
$\beta$ models --- even flattened $\beta$--models with axial ratios within
the observed range (\cite{mohr95}) --- fitting to azimuthally averaged 
surface brightness profiles recovers unbiased estimators of $\beta$ and $R_c$
as long as the point source response function (PSF) of the imager
is properly treated.

\myputfigure{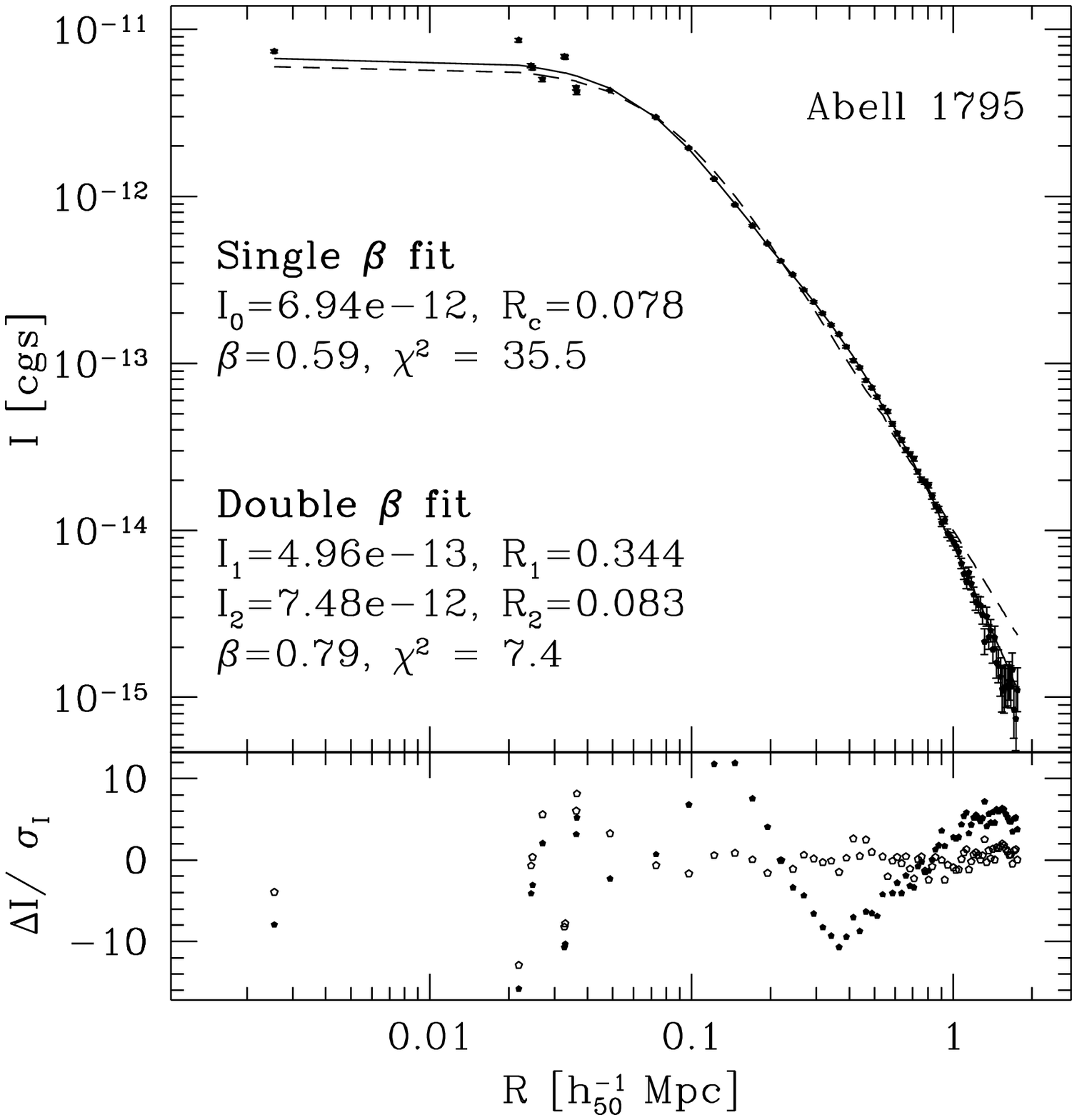}{-3.0}{0.50}{-30}\figcaption{
In the upper panel we plot the observed surface
brightness profile in A\,1795 (points with 
error bars), the best fit single $\beta$ model (dashed line), and the best
fit double $\beta$ model (solid line).  The lower panel contains residuals
around the single $\beta$ (solid) and the double $\beta$ (open) fits.
In clusters with central emission excesses, a single $\beta$ fit
produces an artificially small core radius and $\beta$; adding the second
$\beta$ component results in an improved fit to the emission excess and
the outer profile.
}
\label{doublesinglefit}
\medskip

Unfortunately, approximately half of nearby
galaxy clusters exhibit X--ray emission which is not azimuthally symmetric
(\cite{mohr93,mohr95}).  These features include large centroid variations
(e.g. A\,754) and bimodality (e.g. A\,85) which violate the assumptions
underlying Eqn. \ref{beta}.
Nevertheless, we find that the $\beta$ model describes azimuthally averaged
cluster emission reasonably well and produces sensible results
for most clusters.  As discussed in detail below ($\S3.6$), we 
quantify the systematics associated with $\beta$ model fitting to 
radial profiles by analysing an ensemble of simulated clusters that 
exhibit X--ray morphologies similar to those of our PSPC cluster sample.

The presence of cooling flows presents another potential problem to our
choice of model.  Significant central emission excesses associated with
cooling instabilities are found in 18 of the clusters in our sample;  they
can greatly bias the model parameters.
Clusters for which cooling flows present a problem to fitting are easily
recognized by two criteria. First, the cluster must display nonrandom
behaviour in the fit residuals consistent with a central emission excess.
Second, the cluster must appear ``relaxed'', lacking obvious asphericity
or substructure which would indicate a recent merger.
Figure~\ref{doublesinglefit} shows an example of
these criteria in Abell 1795, a symmetric cluster
which displays a significant central emission excess.  The emission excess
biases our best-fit $\beta$ model (the dashed line) towards a small
core radius and shallow profile, resulting in poor agreement between fit
and data in the region outside the excess.

In such cases we model the emission excess with a second $\beta$ model
and simply fit the cluster surface brightness profile to the sum of
the two:

\begin{equation}
I_D(R) = \sum_{i=1}^2 I_i 
\left[1 + {\left(R\over R_i\right)}^2\right]^{-3\beta+{1\over2}}.
\label{doublebeta}
\end{equation}

We constrain both components of this function to have the same $\beta$,
and we determine the distribution of the underlying ICM numerically. This
method essentially decouples the inner and outer regions of the cluster,
allowing the fit to find the transition between excess and primary
emission on its own. An example of a two-component fit is also shown in
Figure~\ref{doublesinglefit} (solid line); the extra degrees of freedom
are effective in removing nonrandom trends from the residuals. 
In the process of formulating our method, 
we also considered removing the emission excess
and fitting a single component $\beta$ model to the
remainder of the image. 
We found that fitting under these conditions produces biased parameter
estimates in cases where the emission model {\it is} a perfect $\beta$ model;
specifically, the best fit core radius is correlated with the size of
the excised region. These systematics and
associated signal to noise reductions led us to abandon this approach.
The double $\beta$ model approach is similar in spirit to this
method and usually gives similar results, but is much more objective.

\subsection{Calculating Radial Profiles}

Producing a background subtracted cluster surface brightness profile
appropriate for fitting requires several steps.  First, we choose an emission
center using a circular aperture of radius 10 pixels.  We adjust the position
of this aperture to minimize the difference between the geometric center and
the centroid of the X-ray photon distribution which lies within the aperture.
This approach converges even in cases where the cluster emission is 
significantly skew (\cite{mohr93}).  

Second, we evaluate the background level in the image by examining 
annuli at large radii where the cluster surface brightness makes a 
negligible contribution.  Specifically, we measure the background 
surface brightness within 5 concentric, overlapping annuli with 20 pixel 
width and inner radii ranging from 140 to 180 pixels.  The background 
value for each annulus is the mean of the clipped distribution of the 
$\sim2\times10^4$ pixels within the annulus.  The clipping algorithm 
excludes all pixels that are more than 2$\sigma_i$ brighter than the 
median value, where $\sigma_i$ is the uncertainty associated
with pixel $i$.  The clipping algorithm is iterated 10 times, with upper
and lower sigma clipping during the last two iterations.  Tests on
simulated PSPC images indicate that the mean value is an unbiased
estimator of the image background.  

This method allows us to detect clusters whose emission
contaminates the entire image, by seeing whether the background values
steadily decrease with increasing radius.  It also guards against the
possibility of localized excess emission contaminating our result. In cases
where the neighboring clusters contaminate the image background region in one
direction (e.g. A\,3558), we use half rather than full annuli.
In the end, we measure the background in the annulus whose background value
lies closest to the mean of the five measurements; in cases where detectable
cluster emission essentially fills the PSPC FOV, we use the background value
in the outermost annulus.  We use the scatter in the five background
measurements as an indicator of the true background uncertainty.

Lastly, we calculate the radial profile.  Individual pixels are treated
separately within 1.5 pixels of the image centroid;
outside this region the profile values are averages over all pixels 
whose centers lie
within an annulus of 1 pixel width.  The profile is
truncated at the radius $\theta_{max}$
where the signal to noise falls below a critical 
value; specifically, the profile is terminated when the average signal to
noise of the last few profile points falls below $\sim3$.  A few clusters are
treated differently because of contamination by neighboring clusters;
for example, Abell\,401 is fit out to a radius halfway between it
and Abell\,399.

\subsection{Fitting $\beta$ Models}

We find the best estimates of the parameters $I_0$, $\beta$, and $R_c$
by minimizing the $\chi^2$ difference between the PSF convolved model
and the observed surface brightness profile. 
We minimize $\chi^2$ using the downhill simplex method (\cite{press92}).
In correcting for the
effects of the PSF and finite pixel scale we employ the techniques used
in studies of the fundamental plane of elliptical galaxies 
(\cite{saglia93,mohr97b}).  The PSF convolved
mean surface brightness $I_c$ within an annulus of radius $R$ and width $2dR$ is
\begin{equation}
I_c(R)={[F(R+dR)-F(R-dR)]\over4\pi RdR}
\label{betaconv}
\end{equation}
where $F(R)$ is the integrated flux contained within a radius $R$.  The
integrated flux is a function of the $\beta$ model parameters and the PSF.
\begin{equation}
F(R) = R\int_0^\infty\, dk J_1(kR)\hat p(k) \hat I(k)
\end{equation}
where $J_1$ is the Bessel function and $\hat p(k)$ and $\hat I(k)$
are the Fourier transforms of the PSF and the one or two component 
$\beta$ model.  We use a
PSF of the form
\begin{equation}
\hat p(k) = e^{-(kb)^\gamma}
\label{PSFequation}
\end{equation}
where the two free parameters $b$ and $\gamma$ describe the scale and
shape of the PSF.  We determine the best fit PSF parameters $b=0.47$
and $\gamma=1.89$ by fitting to artificial point sources created with
the PROS task {\it rosprf} at 1.0\,keV (see Fig. \ref{PSFfigure});
these parameters
are appropriate for the central part of the PSPC field,
consistent with the position of cluster cores in this dataset.  

\myputfigure{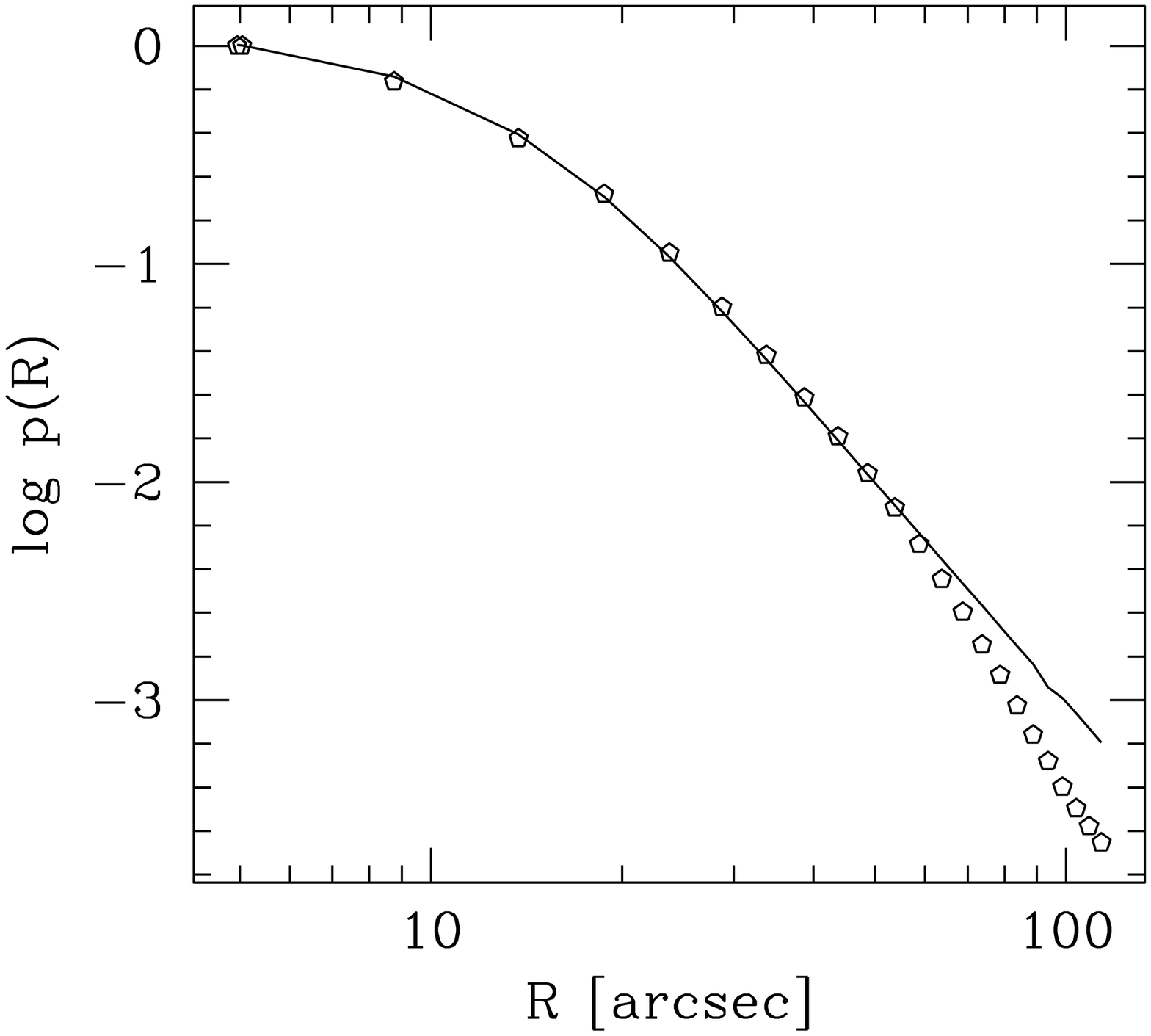}{-3.4}{0.50}{-45}\figcaption{
The radial profile (points; produced using PROS) and fit
(solid line; see Eqn. \ref{PSFequation})
of a 1~keV point source located in the center of the ROSAT PSPC field.
The divergence of the model and the true PSF at radii greater
than 60$''$ does not bias estimates of $\beta$ model parameters.}
\label{PSFfigure}
\medskip

The Fourier transform of the one component $\beta$ model $I(R)$ is
\begin{equation}
\hat I(k)=2\pi I_0 R_c^2 \int_0^\infty\, d\lambda\,
{\lambda J_0(kR_c\lambda)\over\left(1+\lambda^2\right)^{-3\beta+{1\over2}}}
\end{equation}
where $J_0$ is the Bessel function.  This function $\hat I(k)$ is
not analytic for all $\beta$, but note that the integral is only a function
of $\beta$ and the combination $kR_c$.  Our approach is to calculate
$\hat I(k)$ over a finely sampled grid in $\beta$ and $kR_c$ and then
interpolate on this grid when evaluating the PSF convolved surface brightness
$I_c(R)$ (Eqn. \ref{betaconv}).  We test the accuracy of this grid by
comparing the inverse transform of $\hat I(k)$ directly with $I(R)$.
The Fourier transform of the two component $\beta$ model is simply the
sum of the transforms of the individual components.
Naturally, the PSF corrected surface
brightness $I_c(R)$ approaches the uncorrected profile $I(R)$ at radii
which are several times the PSF scale, so at large radii we use the
unconvolved surface brightness $I(R)$ (Eqn. \ref{beta}) directly.

\subsection{Calculating ICM Masses}

We constrain the ICM central density using the cluster
X--ray luminosity measured within an annulus (e.g. \cite{david90,mohr96}).  
Specifically, the luminosity within an annulus defined by a minimum
radius $R_{-}$ and a maximum radius $R_{+}$ is
\begin{equation}
L_X(R_{-},R_{+}) = 4\pi \int_0^\infty dz \int_{R_{-}}^{R_{+}}
n_e(r) n_H(r) \Lambda(T) RdR
\label{luminosity}
\end{equation}
where $n_e(r)$ and $n_H(r)$
are the electron and proton number densities,
$\Lambda(T)$ is the radiative cooling coefficient and $r^2=R^2+z^2$.
The integral along $z$ can be
truncated at the cluster ``boundary''; tests demonstrate that
for observed values of $\beta$ this integral is not sensitive to its upper
limit.  Using $n_e(r)=\rho(r)/\mu_e m_p$ and $n_H(r)=\rho(r)/\mu_H m_p$
the ICM central density is
\begin{equation}
\rho_0=\left({L_X(R_{-},R_{+}) \mu_e \mu_H m_p^2 (1-3\beta)\over 2\pi
\Lambda(\left < T_X\right>) R_c^3 F(R_{-},R_{+})}\right)^{1\over2}
\label{centralrho}
\end{equation}
where $F(R_{-},R_{+})$ is the dimensionless integral
\begin{eqnarray}
F(R_{-},R_{+})= \int_0^\infty ds
\left[\left(1+s^2+({R_{+}/R_c})^2\right)^{1-3\beta}\right. \\
- \left.\left(1+s^2+({R_{-}/R_c})^2\right)^{1-3\beta}\right],
\nonumber
\end{eqnarray}
and we have assumed that the ICM is isothermal with temperature \TX.
For the results
presented here we measure the luminosity over the same region we
fit the X-ray surface brightness profile ($R_{-}=0$ and $R_{+}=\theta_{max}$).

\myputfigure{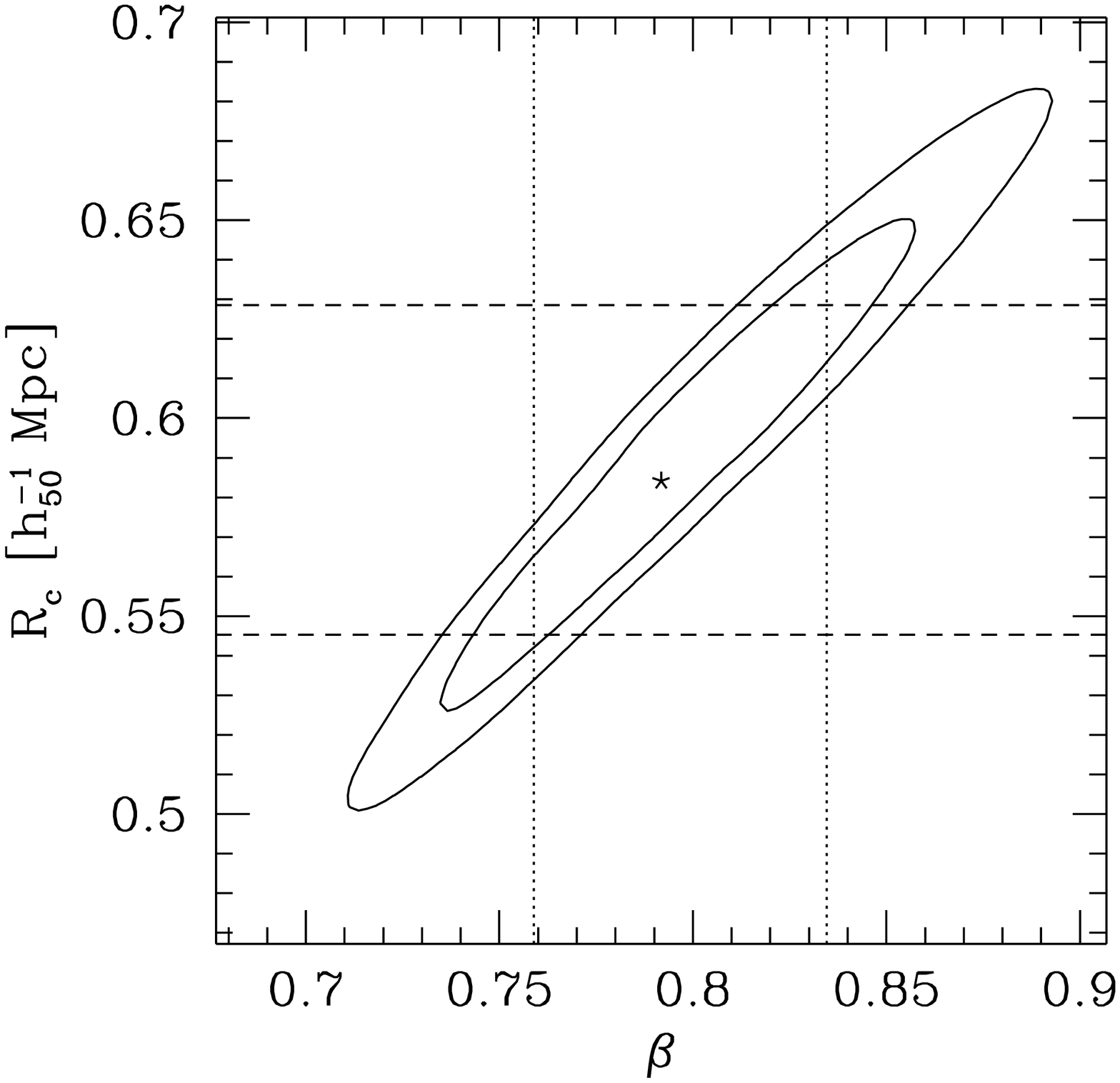}{-3.5}{0.50}{-30}\figcaption{
The correlated errors in fit parameters $\beta$ and $R_c$ are
obvious in this $\chi^2_\nu$ contour plot.  The best fit value
for A\,2255 is marked with a star, the contours deliniate 2$\sigma$ and
3$\sigma$ confidence regions, and the dotted and dashed lines
correspond to 90\% confidence intervals determined by Monte Carlo runs.
The uncertainties presented with tabulated results are derived from
Monte Carlo simulations.
}\label{corrfits}
\medskip

In determining $\rho_0$ we use $\mu_e=1.167$ and
$\mu_H=1.400$, appropriate for a fully ionized plasma with 30\% solar
abundances (\cite{feldman92}).
The ICM mass within some limiting radius is then a straightforward integral
over the $\beta$ model density distribution using the measured
ICM central density (Eqn. \ref{centralrho}).

The density distribution consistent with a two component $\beta$ model
is not analytic in general.  We express the density distribution in the
two component case as a radially dependendent multiple $f(r)$ of the broad,
primary component: $\rho_D(r)=f(r)\rho_1(r)$;  we solve for $f(r)$ using
the surface brightness profiles of the two component fit $I_D(R)$
as a constraint:
\begin{equation}
I_D(R) = 2 {\Lambda(\left<T\right>)\over\mu_e\mu_H m_p^2}
\int_R^\infty {rdr\over\sqrt{r^2-R^2}} 
f^2(r)\rho_1^2(r)
\end{equation}
Using this equation, we work our way inward
solving for $f(r)$ at equal intervals in $r$.  The central density
in this case is $\left.\rho_0=f(r)\rho_1(r)\right|_{r=0}$.
In our sample the central value of the multiple $f(r=0)$ ranges
from 2.5 (Ophiuchus) to 33 (A\,2204) and averages $\sim$10.

\subsection{Measurement Uncertainties}

A uniform, systematic analysis of the ICM properties in this well
defined cluster sample requires careful quantification of all known
sources of uncertainty.
We identify multiple sources of uncertainty in the measured properties 
of the ICM, including Poisson noise, choice of cluster emission center,
PSF blurring, X--ray background subtraction,
ICM temperature uncertainties, luminosity uncertainties, 
and $\left<T_X\right><1.5$~keV gas associated with central 
cooling instabilities.
Below we discuss each in turn, and describe how they are incorporated into
the final uncertainties for the $\beta$ model parameters and ICM masses
\MICM.  We also quantify the impact of each error contribution
on the derived ICM mass, to give some indication of their relative
importance.

\begin{itemize}

\item{\it Poisson noise}:  We evaluate the Poisson
contribution to the uncertainties using a Monte Carlo approach. 
Specifically, we create and fit 500 independent
realizations of each cluster image.  The template for these artificial images
is a Gaussian smoothed ($\sigma=30$\arcsec)
version of the original cluster image.
Each pixel in an artificial image contains a value Poisson distributed around
the mean of the pixel value in the template image.  When fitting $\beta$
models to the artificial images, we adjust the PSF parameters to account for
the degraded resolution in the smoothed template. This approach allows us
to account for the deviations of many clusters from idealized $\beta$ models.
We use this ensemble of 500 fits for each cluster to quantify the
Poisson contribution to measurement uncertainties.  The typical
Poisson contribution to the fractional uncertainty in \MICM\ is 1.1\% 
(sample median), and the largest is 3.9\% in A\,1689.

The fit parameters $R_c$ and $\beta$ are strongly correlated.
In Fig. \ref{corrfits} we present  2$\sigma$ and 3$\sigma$
$\chi^2_\nu$ contours in the space of $R_c$ and $\beta$ for the single
component fit to the cluster A\,2255.  Rather than present these plots for
each cluster we note the upper and lower 90\% confidence intervals on
the best fit parameters.  Dashed lines
indicate the locations of these confidence boundaries in Fig. \ref{corrfits}.

\item{\it Emission center}:  We use a circular aperture with 150$''$ radius
to define the cluster emission center (see $\S3.2$)
before calculating the radial profile.
In clusters with azimuthally symmetric emission,
the emission center is independent of the aperture radius.  In clusters that
exhibit centroid variations this is not the case.
Calculating the profile around different
emission centers can affect core radii, $\beta$'s and ICM masses.
We evaluate these effects by
re-analyzing the cluster images using a centering aperture three times larger.
We find that the measured properties of
the ICM are insensitive to centering aperture radius except for those
few clusters with grossly asymmetric X-ray images.  The substructure
in these clusters indicates recent major mergers, and, for the most part,
they can be identified by the high $\chi^2_\nu$ in the fit results.
We use the difference in ICM parameters in the two different analyses
as an estimate of the scale of this uncertainty.  The
ratio of \MICM\ measurements with the two centering apertures
has a mean value of 0.997 and a standard deviation of 2.5\% calculated
over the whole sample; the largest outlier is A\,754, a well studied
recent merger candidate (e.g. \cite{fabricant86,henrik96,roettiger98}),
where $M_{ICM}$ changes by 16\%.

\item{\it PSF correction}:  Without a PSF correction, core radii and $\beta$'s 
in small core radius clusters would be systematically
overestimated.  By fitting PSF corrected $\beta$
models to observed radial profiles, we remove this bias from
our $\beta$ parameters and ICM masses.
We ignore any residual errors associated with
the minor differences between the true PSPC PSF and our best fit model (see
Fig. \ref{PSFfigure}).

\item{\it X--ray background}:  Errors in the background subtraction
can systematically bias $\beta$ parameters and X-ray luminosities.
We use the variation in the measured background within five concentric annuli
as a measure of the true background uncertainty $\sigma_{XB}$ (see primary
discussion in $\S3.1$). We quantify
the effects of the background uncertainty on measured ICM parameters
by refitting the image with the background fixed to the preferred value
perturbed by $\pm1\sigma_{XB}$ and $\pm$2$\sigma_{XB}$. 
For each cluster we use the five fits to measure a maximal deviation
in the parameters of interest (corresponding to a $\sim4\sigma_{XB}$
variation in the X-ray background); we use half this maximal difference as
an estimate of the X-ray background induced uncertainty.
In our ensemble, the RMS maximal deviation in \MICM is 3.6\%, and the largest
change is 14\% in MKW~3S.

\item{\it ICM \TX\ measurements}:
Typical measurement uncertainties in \TX\ translate into 
negligible uncertainties in the ICM parameters, but make important contributions
to uncertainties in binding mass estimates.  Eqn \ref{centralrho} shows that
$\rho_0\propto\sqrt{L_X/\Lambda}$ where $L_X$ is the X-ray luminosity and
$\Lambda$ is the cooling coefficient.  The temperature dependence of $L_X$ and 
$\Lambda$ is approximately the same, because to determine $\L_X$ we
measure the cluster X-ray count rate in the PSPC and
then use an ICM emission model $\Lambda$,
galactic absorption, cluster distance and the PSPC effective area tables to
convert to a bolometric luminosity (PROS tasks). 
We quantify this effect by varying the
temperature of the cluster A\,1795 ($\left< T_X\right>=5.3$\,keV) from
1 to 10\,keV and calculating the resulting variations in $\rho_0$. 
Above 5\,keV $\delta\rho_0\sim\delta T_X^{0.055}$, so a factor of two
increase in \TX\ corresponds to a $\sim4$\% increase in $\rho_0$; 
between 1\,keV and 5\,keV uncertainties in \TX\ have an even smaller 
effect on $\rho_0$.

Temperature uncertainties dominate the formal uncertainties in the binding mass.
We account for these uncertainties in
measurements of \MICM\ and \fICM; note that \MICM\ measured within a 
constant 1$h^{-1}_{50}$~Mpc aperture is unaffected by \TX\ uncertainties.

\item{\it Luminosity measurements}:  We measure cluster aperture
luminosities using the PSPC images. 
With Eqn. \ref{centralrho} and the uncertainty
in the observed cluster count rate, we calculate the contribution to
uncertainties in the ICM parameters.  The typical contribution to
\MICM\ uncertainties is 0.4\% (sample median) and the largest contribution
is 1\% in A\,2244.  We do not account for
any errors in the PSPC effective area tables; upcoming observations
with AXAF will provide an excellent test of these data.
 
\item{\it Core ICM with $T_X<1.5$~keV}: 
The PSPC count rate for a parcel of gas with constant emission measure
is relatively insensitive to $T_X$ for
$T_X>1.5$~keV (see Fig. \ref{PSPCemis}); therefore,
our assumption of isothermality when we calculate the
underlying density profile consistent with the X-ray surface brightness
profile does not affect our results.  However, ICM temperatures within the
cores of clusters which exhibit strong emission excesses
or cooling flows can be less than 1.5~keV.  In clusters where this
is the case the ICM in the
cores is essentially overrepresented with photons
in the cluster X-ray image.  Specifically, the PSPC count
rate for a 1~keV (0.75~keV) gas clump is $\sim$1.4 ($\sim2$) times
higher than the count rate for a 5~keV gas clump.

In the absence of detailed ICM temperature profiles, we
estimate the effects of these cool cores empirically.
We use the clusters with two component $\beta$ model
fits to estimate the nature of this effect on \MICM\ measurements. 
Specifically, we assume that the secondary component of the fit is
responding to ICM at a temperature of 0.75~keV; thus, we
scale the best fit $I_2$ by ${1\over2}$  to
account for the factor of two
higher PSPC count rate, and then recalculate the
ICM distribution and \MICM, taking care to account for the change
in the cluster aperture luminosity.  Results indicate that our
standard isothermal analysis overestimates the true \MICM\ in this
case; in our sample of 18 clusters with double $\beta$ model fits, \MICM\ is
overestimated by $\sim3.5$\% on average,
with a maximal overestimate of 14\% in A\,2204.  The ICM radial distribution
is less affected: \rave\ is underestimated by less than 1\%.

We emphasize that the toy model presented here
overestimates the effects of cool gas on \MICM\ measurements,
because we have assumed that cool gas extends throughout the cluster
(traced by the secondary $\beta$ component), whereas the cool gas is
likely confined to the core regions
where the cooling time falls well below the Hubble time.

\end{itemize}

We combine these sources of uncertainty as follows.  First,
these different error terms are independent of one another and we 
assume that they are each Gaussian distributed; we combine
all but the Poisson contribution in quadrature and call it $\sigma_{NP}$.
Second, we convolve the 500 Monte Carlo results for each cluster with
a Gaussian of scale $\sigma_{NP}$, and
integrate over the distribution to calculate the 90\% confidence intervals.
These confidence intervals appear with the results.

\myputfigure{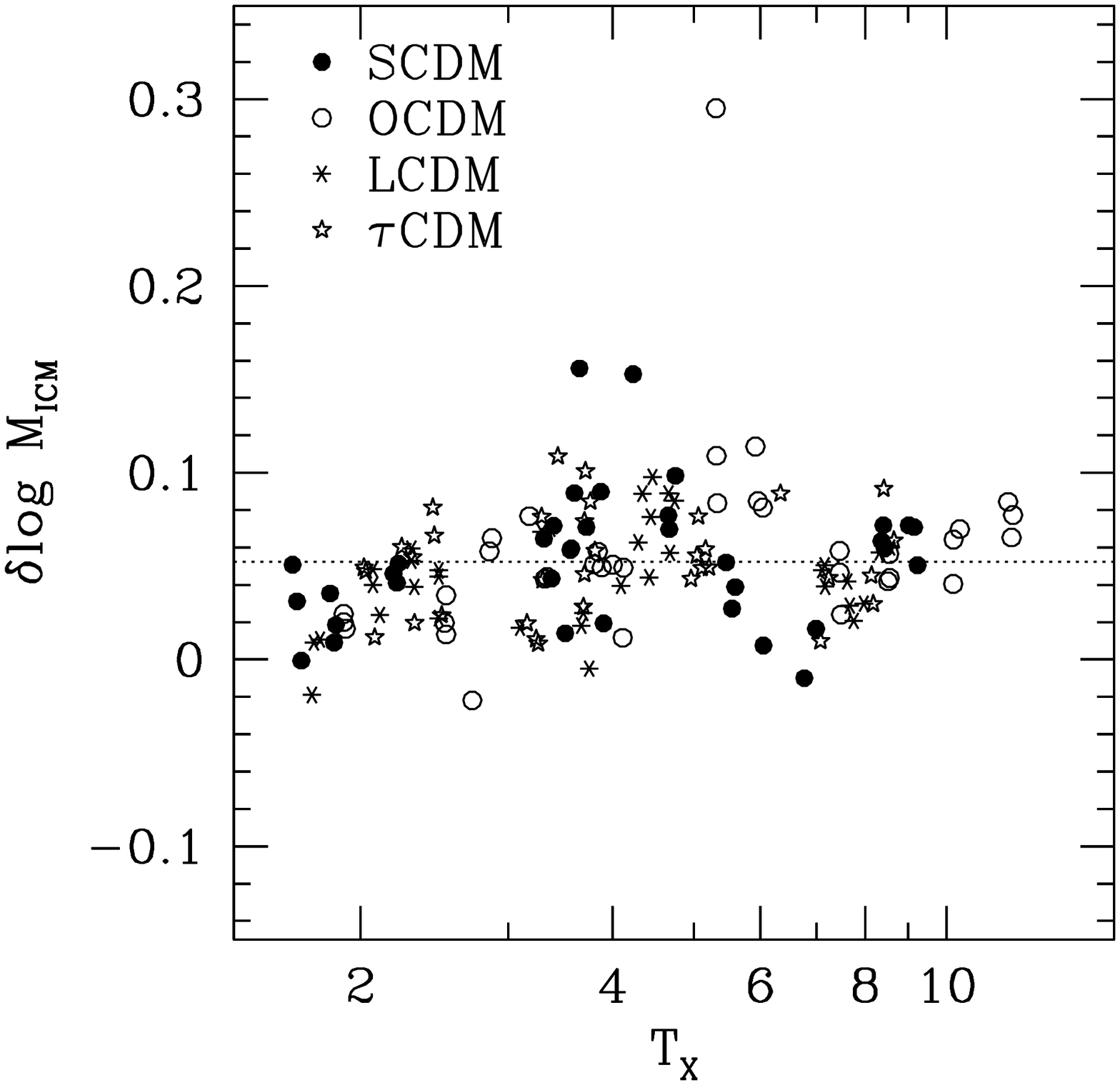}{-3.5}{0.50}{-30}\figcaption{
We plot the log of the ratio of the inferred to the true \MICM\
($\delta \log{M_{ICM}}$) versus \TX for the ensemble of hydrodynamic
simulations.   Estimates of \MICM\ lie systematically higher than
the true values;  the dotted line shows the median overestimate of 12\% 
for the set of 144 images.
}\label{ICMsimulations}
\medskip
\subsection{Testing Our Technique with Simulated Clusters}
We test the accuracy of our ICM measurements using an ensemble
of 48 hydrodynamical cluster simulations (discussed in $\S$2.3). 
The simulations allow us to examine the effects of (i) departures of
the true ICM density distribution from the $\beta$ model and (ii)
ICM temperature variations within the cluster.
The simulated clusters experience present epoch mergers consistent with
expectations for their CDM cosmogonies;
X-ray morphologies of the simulated clusters
in three of the four cosmological models are statistically consistent with the 
X-ray morphologies of the PSPC cluster sample, with the \tCDM\ simulations
exhibiting somewhat more substructure than observed clusters
(Mohr \& Evrard, in prep).  We analyze images of these simulated 
clusters as though they were true cluster images and then compare 
the estimates of the ICM mass fraction to the real values.

Interestingly, we overestimate the true ICM masses; the median overestimate
for a sample of 144 images is 12\%, and the RMS scatter around this median
is $\sim9$\% (see Fig. \ref{ICMsimulations}).  This bias arises 
because the ICM distribution in simulated clusters is not perfectly 
described by an isotropic, smooth density function.  ICM density 
fluctuations around the mean value at a given radius enhance the 
X-ray emissivity of the gas which is proportional to the square of the
density.  We examine the ICM distribution
in the simulations and measure the quantity $\delta_{ICM}=\left<\rho\right>/
\sqrt{\left<\rho^2\right>} -1$ where $\rho$ is the local ICM density; 
because the ICM emissivity $\epsilon_X$ scales as $\epsilon_X\propto\rho^2$,
the overestimate in \MICM\ scales linearly with $\delta_{ICM}$.
We find that within $r_{500}$ the fluctuation
amplitude is $\delta_{ICM}\sim13$\%, in good agreement with the median
overestimate in \MICM.  

These density fluctuations are presumably residual structure in the
ICM from recent mergers; higher resolution simulations would likely 
yield even higher fluctuation amplitudes.  We are carrying out a detailed
analysis of the ICM structure in our simulated ensemble to better characterize
the source of these density fluctuations (Mathiesen, Evrard \& Mohr in prep).
In real clusters this mechanism and additional physics like magnetic 
fields and radiative cooling instabilities (all explicitly ignored in 
these simulations) could all contribute to ICM density fluctuations 
(and overestimates of \MICM).  Quantifying the \MICM\ measurement biases 
in real clusters requires detailed analysis of high resolution,
spatially resolved, X-ray spectra of the ICM emission or comparison
of \MICM\ measurements from X-ray data and observations
of the ICM's Sunyaev-Zeldovich effect on the cosmic microwave background.

In summary, our ensemble of hydrodynamical cluster simulations indicates
that (i) our analysis technique is likely to overestimate \MICM,
and (ii) (after accounting for this bias) high signal to
noise X-ray images of a coeval cluster population
allow the determination of \MICM\ with a typical accuracy of $\sim9$\%.

\section{RESULTS}

Below we present results of an analysis of our X-ray flux
limited sample of 45 galaxy clusters.  First we describe our 
analysis of the cluster X-ray surface brightness profiles.
Second, we present our constraints on ICM masses \MICM\ and mass fractions
\fICM, and third, we constrain systematic variations in the radial 
distribution of the ICM.  Finally, we compare our results to previously
published analyses.

\subsection{X-ray Surface Brightness Profiles}

Best fit parameters for the profiles
appear in Table \ref{SBresults}. Columns contain
the cluster name, the central surface brightness and core radius 
of the primary $\beta$ model component, the central surface
brightness and core radius of the secondary component (if there is one),
the best fit $\beta$, the reduced $\chi^2_\nu$, the number
of constraints in the fit, and the truncation radius of the fit $\theta_m$ 
(we also measure our aperture luminosities within $\theta_{m}$).
Central surface brightnesses are the unabsorbed values in
ergs/s/cm$^2$/\sq\arcmin\ for the 0.5-2.0~keV band. 
Core radii are in $h^{-1}_{50}$~Mpc
(with $q_0=0.5$), and the truncation radii $\theta_m$ are in arcminutes. 
Uncertainties are 90\% confidence intervals with the other fit parameters
unconstrained.

\begin{figure*}
\plotfiddle{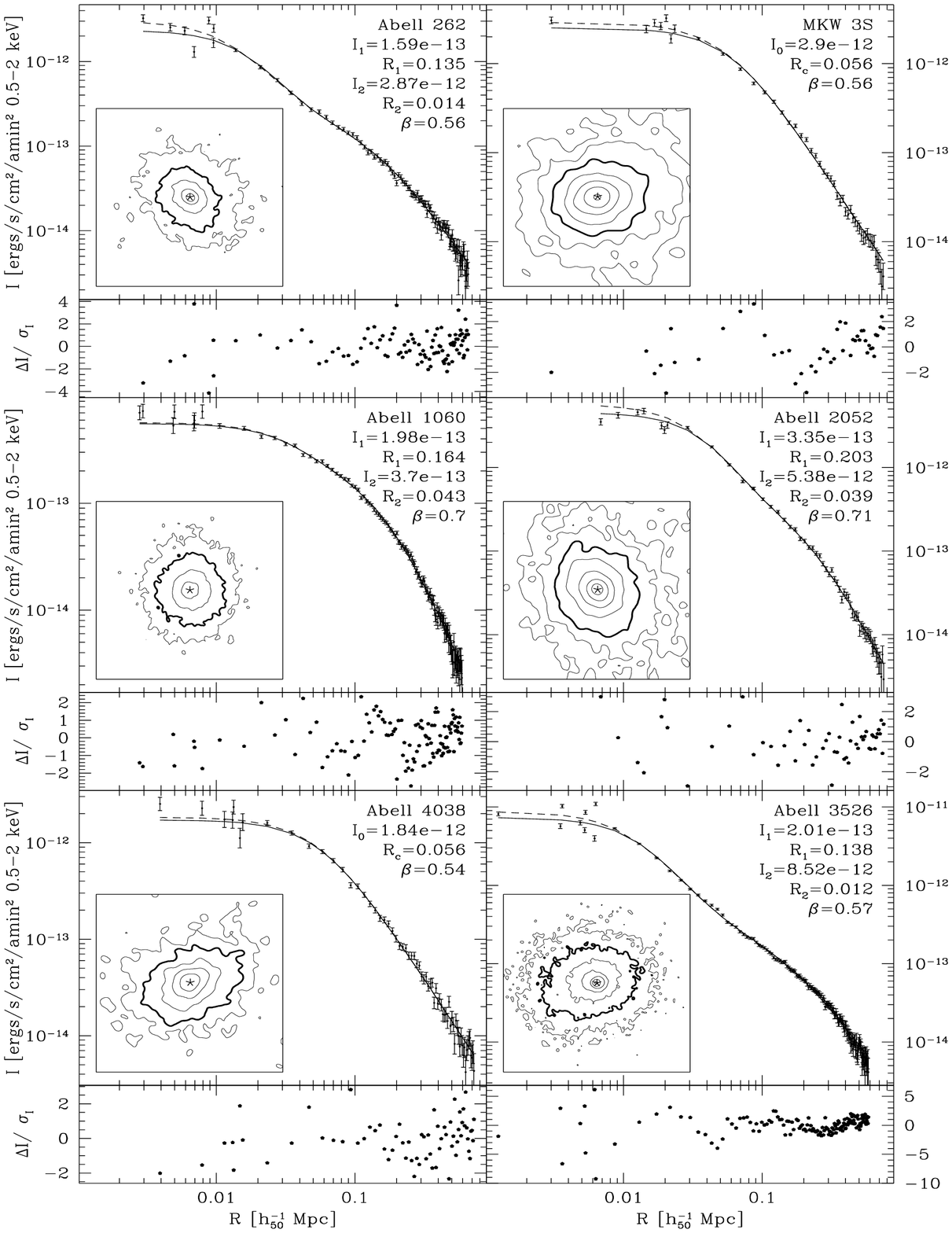}{9.0in}{0}{95}{95}{-290}{-50}
\caption{A\,262, MKW 3S, A\,1060, A\,2052, A\,4038, A\,3526}
\label{profiles}
\end{figure*}

The X-ray surface brightness profile of each cluster appears in
Fig. \ref{profiles}.  Each page contains a two by three
arrangement of cluster panels.
The clusters are arranged by \TX, which facilitates the fixed radial
range for each group of six.  Cluster names appear
in the upper right corner of each panel.  Surface brightness (in
ergs/s/cm$^2$/\sq\arcmin\ in the 0.5-2.0~keV band) is plotted versus
radius in $h^{-1}_{50}$~Mpc.  We plot
measurements (points with error bars), best fit
profile (solid line), and best fit
profile without the PSF correction (dashed line)
for each cluster.  The best fit parameters
appear below the cluster name.  The 
fit residuals are at the bottom of each panel. 
An X-ray contour plot is located at the lower left of each panel;
contours are of a smoothed version of the portion of
the X-ray image used to calculate the
surface brightness profile. Contours appear at
factors of 2.5 in surface brightness,
and the heavy contour always corresponds
to $4\times10^{-14}$~ergs/s/cm$^2$/\sq\arcmin.

We fit two component $\beta$ models to 18 clusters in our sample; 
these clusters exhibit trends in the residuals of their fits to 
single component $\beta$ models.  Some remaining clusters with 
single component $\beta$ fits exhibit trends, but these clusters also 
exhibit large centroid variations or other evidence of recent mergers, and 
so we do not attempt a double component fit: (A\,754, A\,1367, A\,2319,
A\,3266, A\,3558, A\,3667).
Two clusters with double component fits also exhibit obvious evidence of
recent mergers: A\,2142 and A\,85; in the case of A\,85
we remove the small, secondary surface brightness peak to the south before
calculating a radial profile.

It is interesting to note that 29
clusters in our sample are not formally well fit by
either a one or two component $\beta$
model (less than 1\% chance of consistency
between model and data).  This is most likely
another indicator of the prevalence of cluster X-ray
substructure (\cite{mohr95}).  However, in all but a few clusters
the best fit $\beta$ model reproduces the general character of the
surface brightness profile (no systematic trends in the residuals), and, as
demonstrated by our analysis of the cluster simulations,
this is all that is required to constrain the average behavior of the
density profile and calculate \MICM\ with a typical accuracy of $\sim$9\%. 
(Note that parameter uncertainties are determined using Monte Carlo
simulations and not $\Delta\chi^2$.)

\begin{figure*}
\plotfiddle{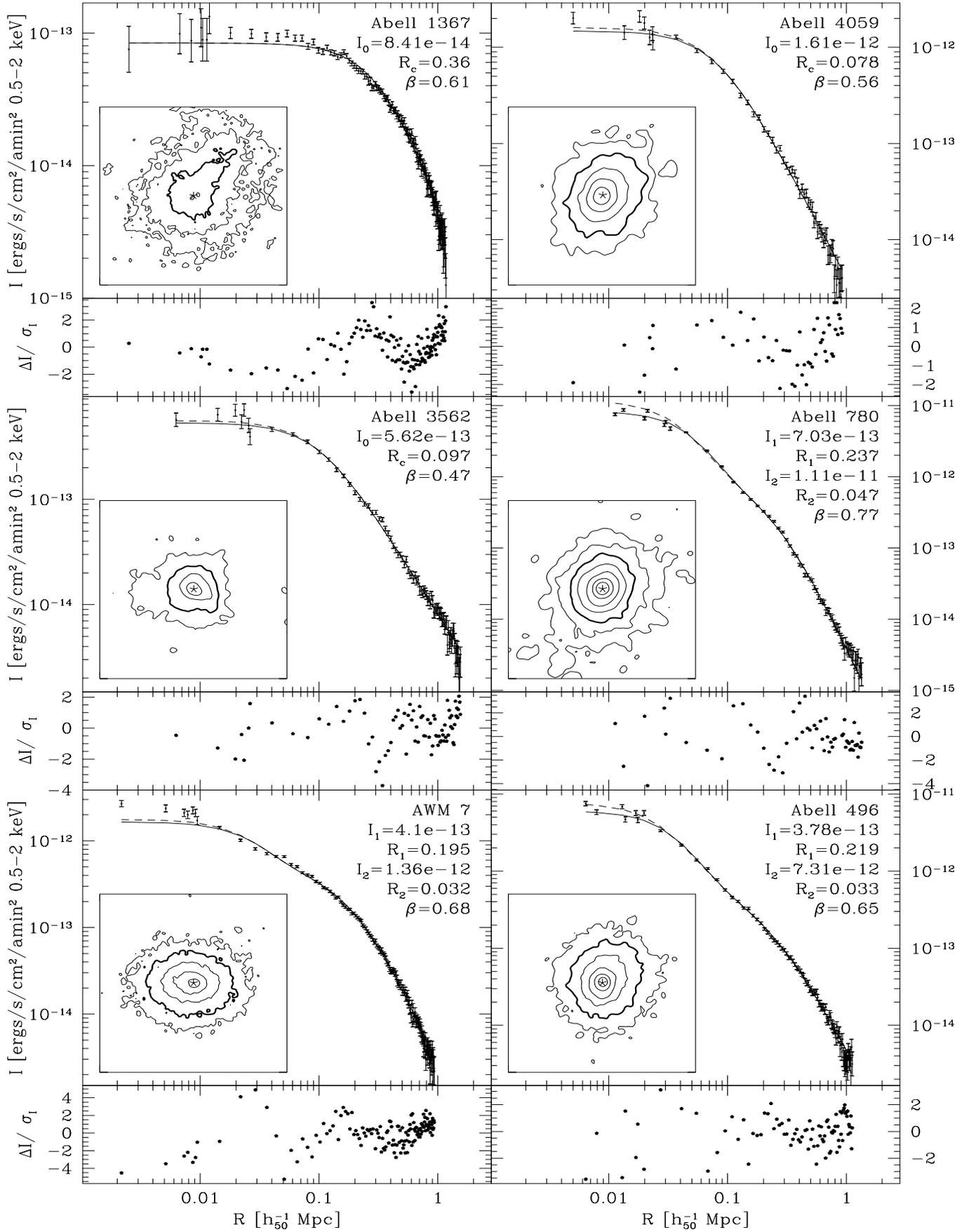}{9.0in}{0}{95}{95}{-290}{-50}
\figurenum{6b}
\caption{A\,1367, A\,4059, A\,3562, A\,780, AWM 7, A\,496}
\end{figure*}

\begin{figure*}
\plotfiddle{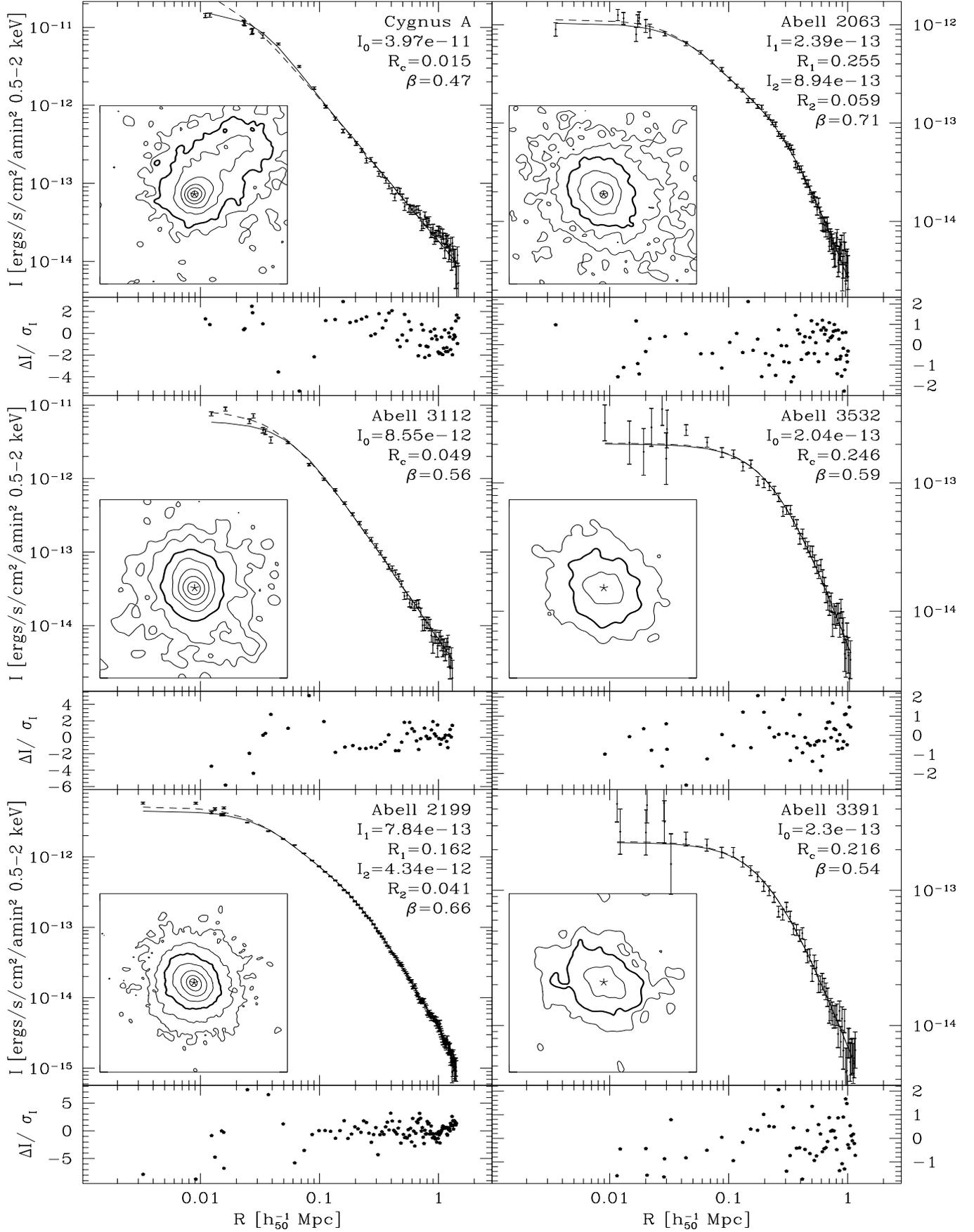}{9.0in}{0}{95}{95}{-290}{-50}
\figurenum{6c}
\caption{Cygnus A, A\,2063, A\,3112, A\,3532, A\,2199, A\,3391}
\end{figure*}

\begin{figure*}
\plotfiddle{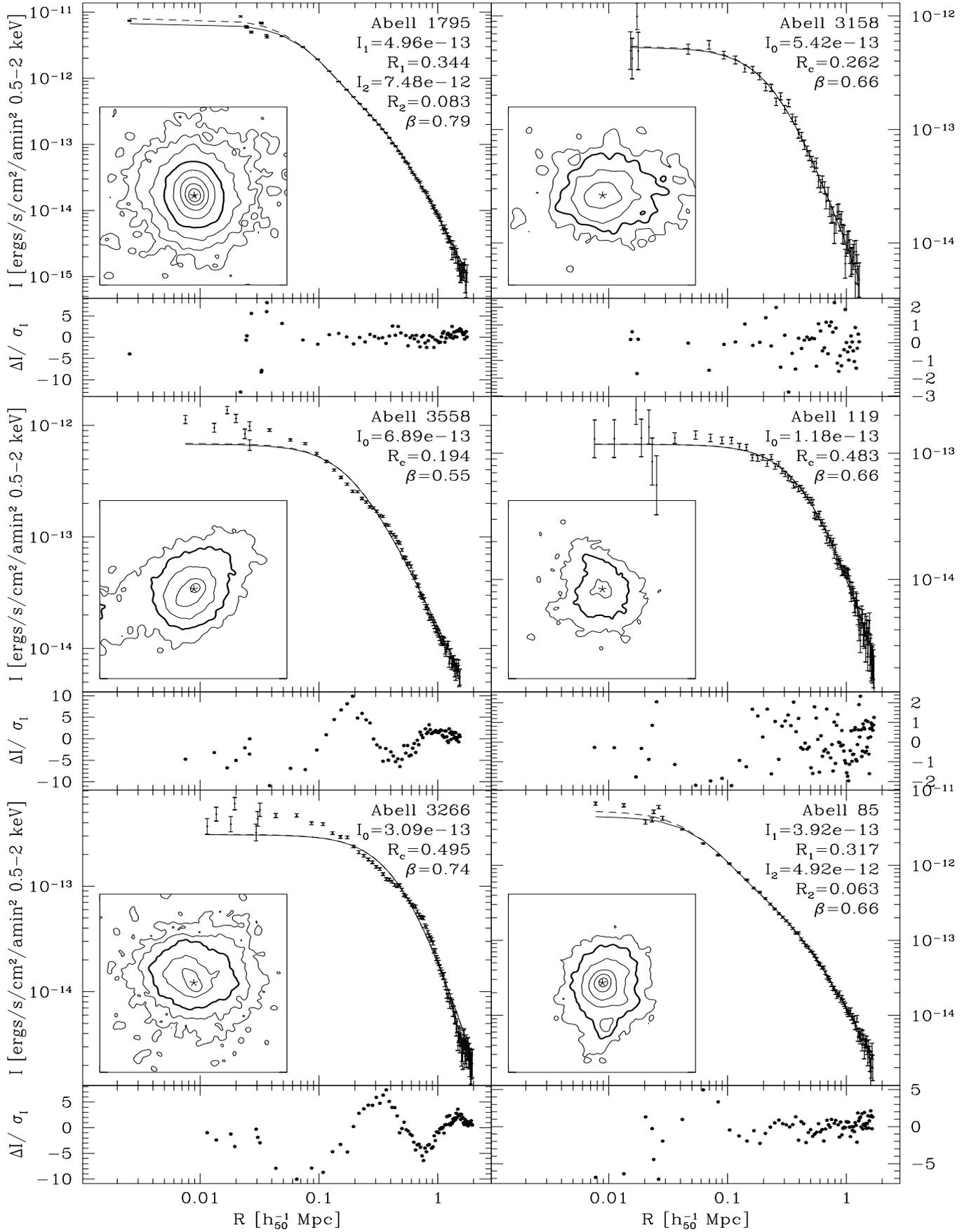}{9.0in}{0}{95}{95}{-290}{-50}
\figurenum{6d}
\caption{A\,1795, A\,3158, A\,3558, A\,119, A\,3266, A\,85}
\end{figure*}

\begin{figure*}
\plotfiddle{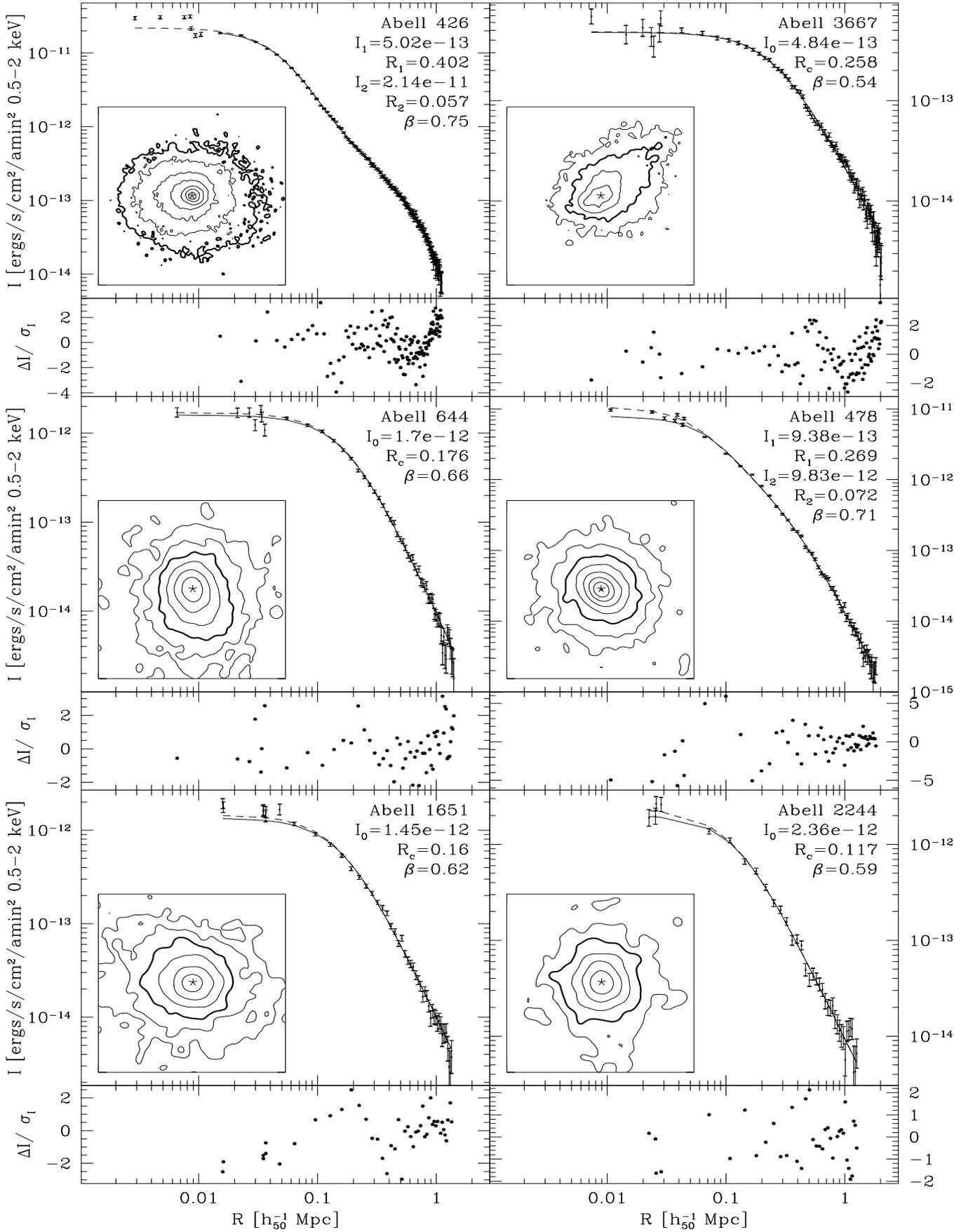}{9.0in}{0}{95}{95}{-290}{-50}
\figurenum{6e}
\caption{A\,426, A\,3667, A\,644, A\,478, A\,1651, A\,2244}
\end{figure*}

\begin{figure*}
\plotfiddle{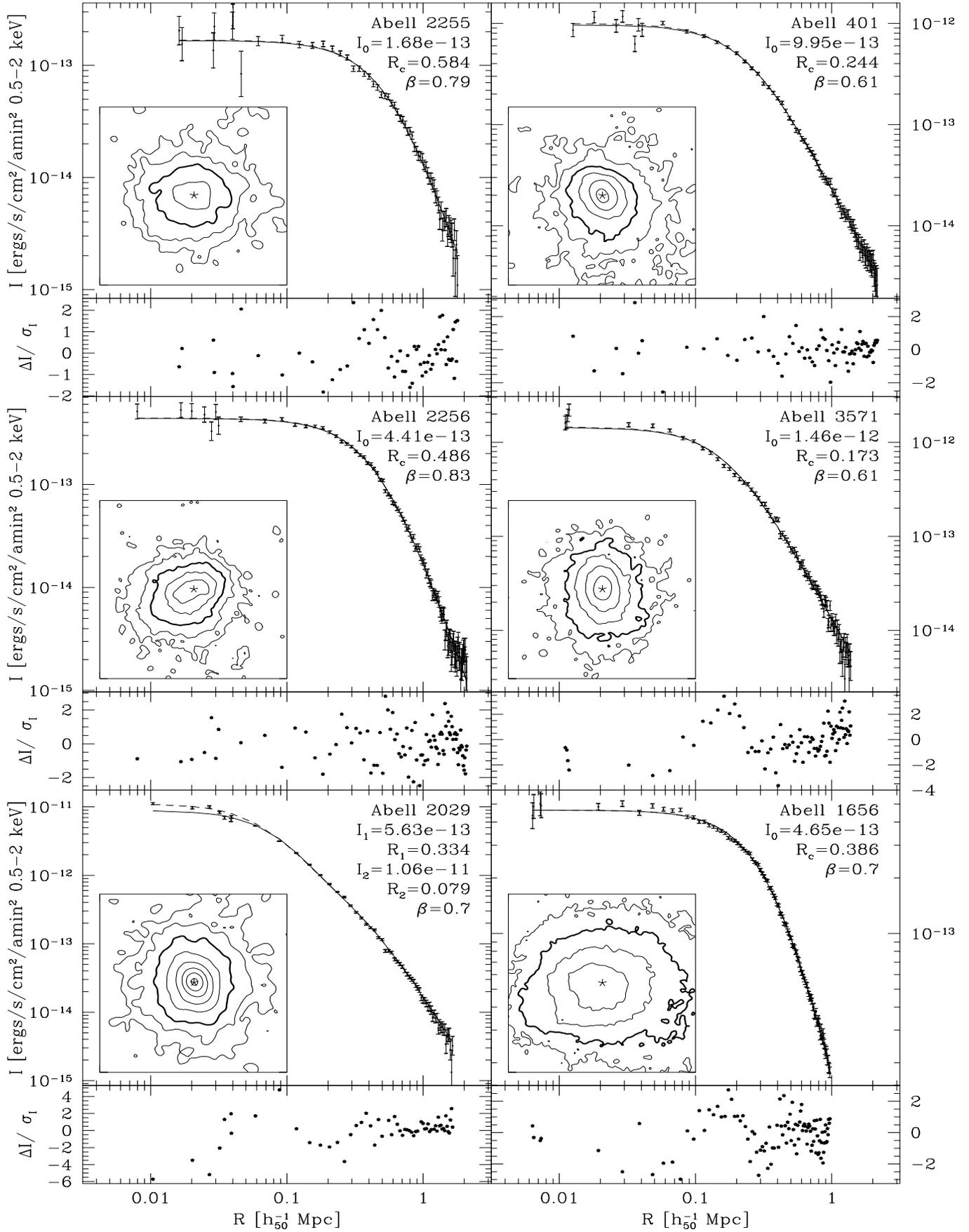}{9.0in}{0}{95}{95}{-290}{-50}
\figurenum{6f}
\caption{A\,2255, A\,399, A\,2256, A\,3571, A\,2029, A\,1656}
\end{figure*}

\begin{figure*}
\plotfiddle{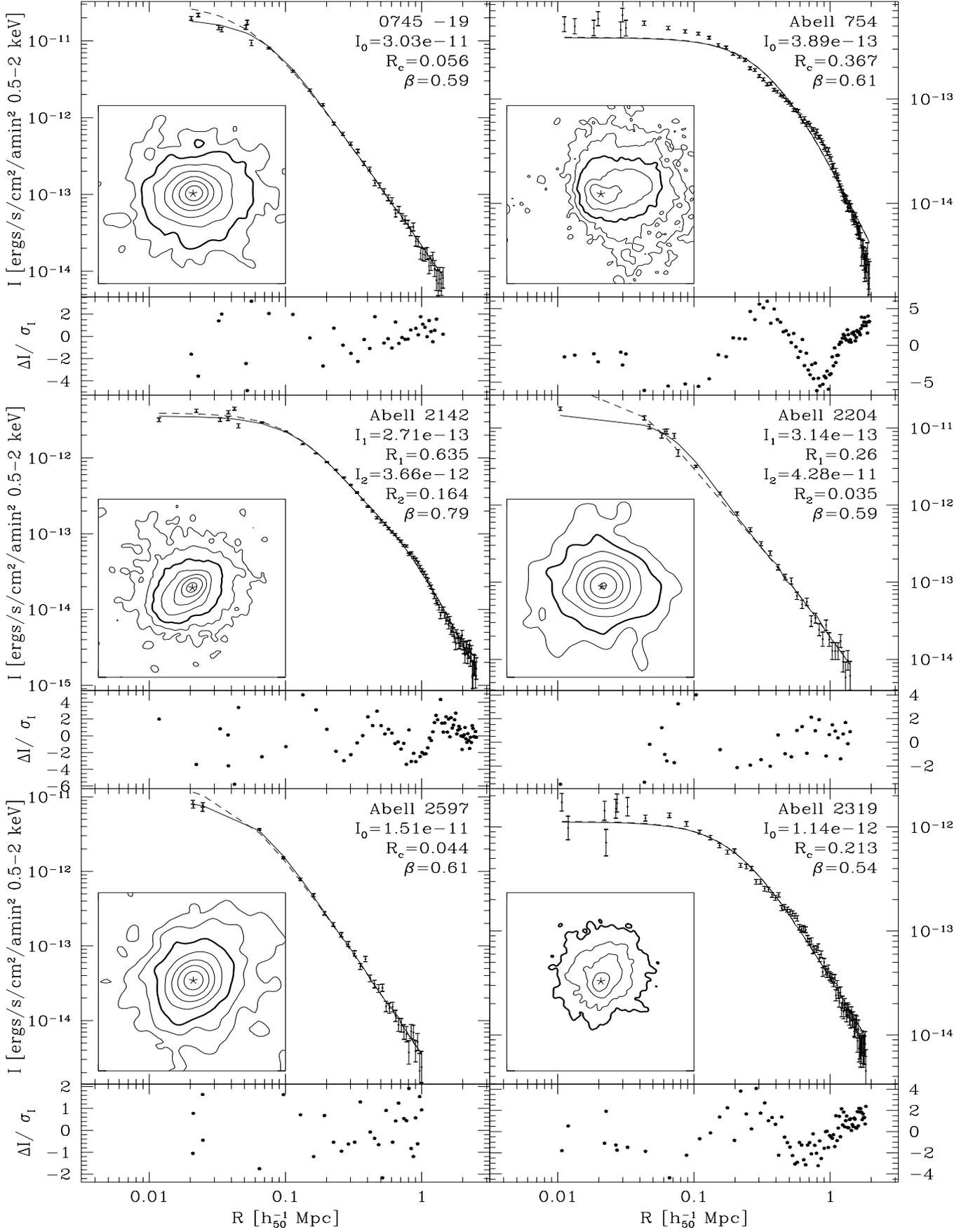}{9.0in}{0}{95}{95}{-290}{-50}
\figurenum{6g}
\caption{0745 -19, A\,754, A\,2142, A\,2204, A\,2597, A\,2319}
\end{figure*}

\begin{figure*}[htb]
\figurenum{6h}
\plotfiddle{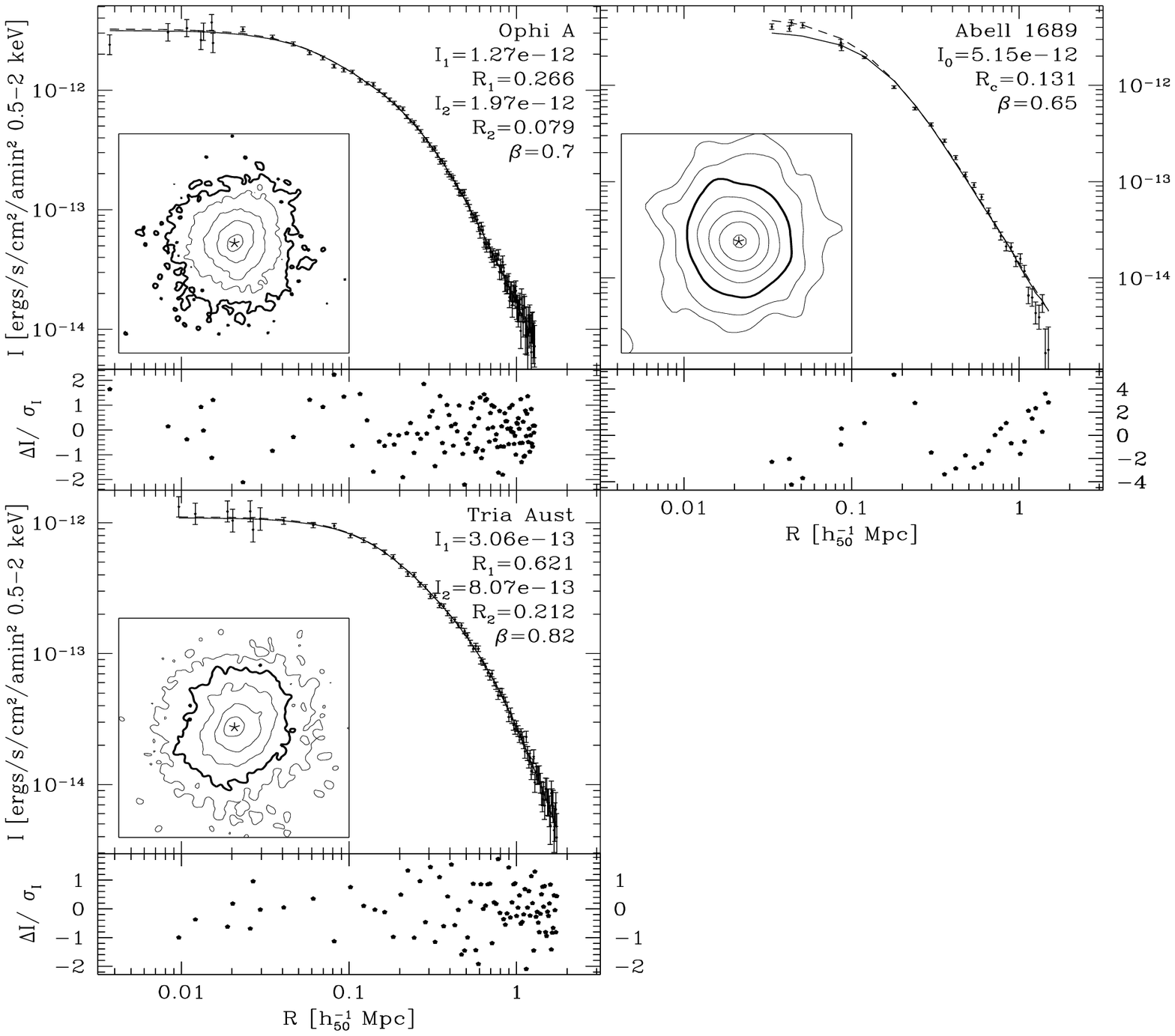}{6.2in}{0}{95}{95}{-290}{-260}
\caption{Ophiucus, A\,1689, Triangulum Australis;  
Cluster X--ray surface brightness profiles:  For each cluster
the radially averaged, unabsorbed surface brightness in units
of ergs/s/cm$^2$/\sq\arcmin in the 0.5:2.0~keV band
is plotted versus distance from the cluster emission center
[in $h^{-1}_{50}$~Mpc].  The surface brightness measurements (points
with error bars), the underlying model (dashed line) and the PSF
corrected model (solid line) are plotted. 
Residuals between the fit and the data
(scaled by the data uncertainty) appear below each radial profile.  A
contour map of the portion of the cluster image used to calculate $I(R)$ 
appears in the lower left corner; the images are smoothed to facilitate
contouring.  The emission center is marked with a star; contours appear
at factors of 2.5 in surface brightness,
and the heavy contour corresponds to a fiducial surface brightness
of $4.0\times10^{-14}$~ergs/s/cm$^2$/\sq\arcmin in each cluster.
The cluster name and best fit parameters appear in the upper right hand
corner of each plot.  Both sets of parameters are listed for those
clusters fit with a two component $\beta$ model.}
\end{figure*}

Note that the high $\chi^2_\nu$ fitting problem for galaxy clusters
is similar to that of elliptical galaxies. Elliptical galaxies
are reasonably well described by bulge plus disk models, but very rarely are
the measured galaxy profiles and models consistent in a formal
$\chi^2$ sense (\cite{saglia97}); interestingly, the photometric 
parameters derived from those fits exhibit the fascinating regularity 
relation termed the Fundamental Plane.

\begin{figure*}[hb]
\plotfiddle{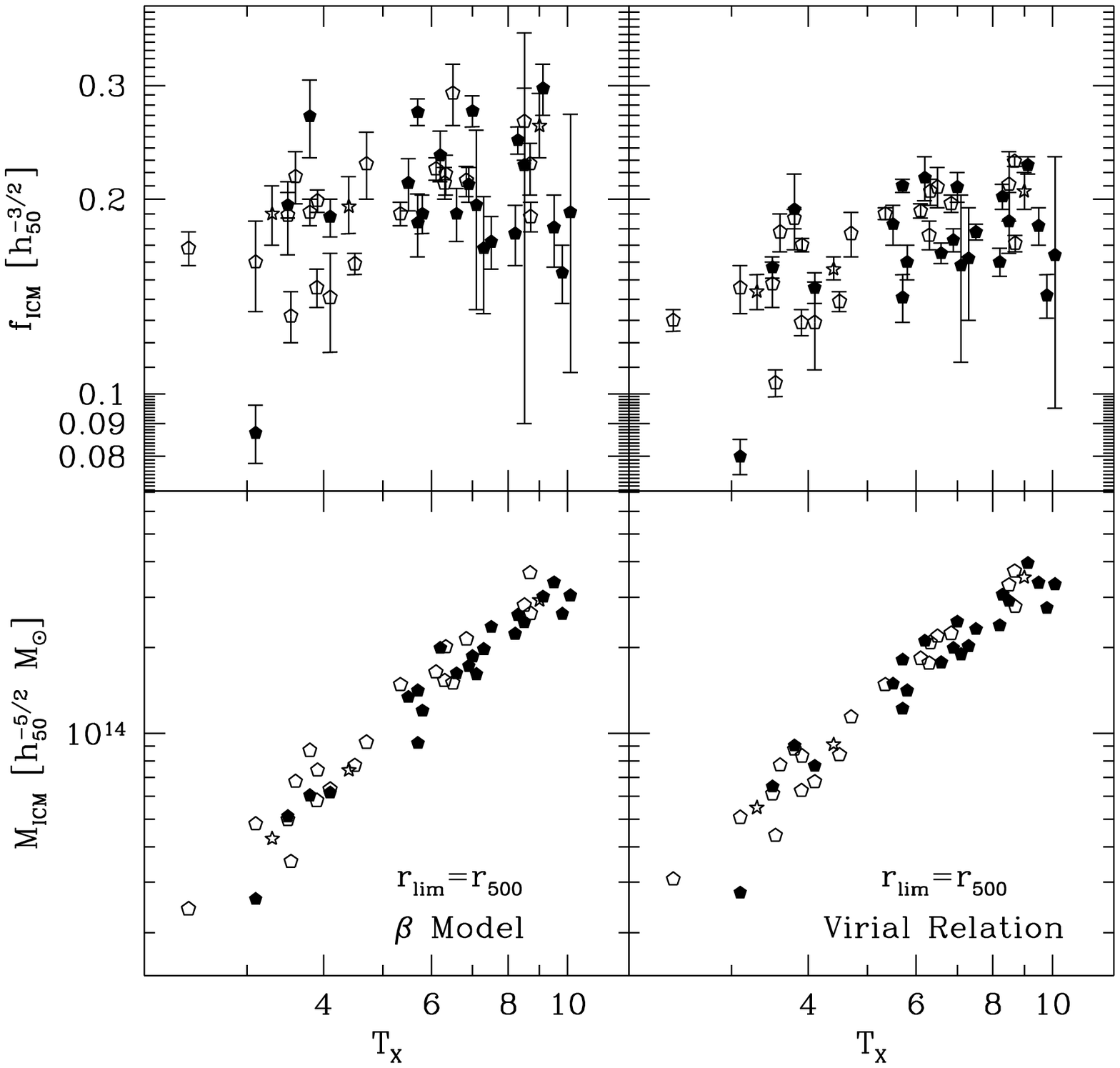}{6.1in}{0}{90}{90}{-280}{-160}
\caption{
We plot ICM mass \MICM\ (below) and mass fraction \fICM\ (above) versus \TX; 
these quantities are measured within the limiting radius $r_{lim}=r_{500}$,
where \rfive\ is calculated using the virial relation (left) and the
isothermal $\beta$ model (right; see $\S$4.2 and Table \ref{ICMresults}). 
Different symbols represent clusters with central cooling times
significantly below 10$^{10}$~yrs (open), higher central cooling times (solid),
and clusters with no upper limits on their \TX\ measurements (stars).
}\label{ICM500}
\end{figure*}

\subsection{ICM Properties}

We use the fit to the X-ray surface brightness profile of each cluster
to constrain the ICM central density, radial distribution, and total mass
within several limiting radii $r_{lim}$.  We use $r_{lim}=1h^{-1}_{50}$~Mpc
and $r_{lim}=r_{500}$, the radius within which the mean density is
500 times the critical density $\rho_{crit}=3H_0^2/8\pi G$.  By using
\rfive\ we are able to study the same portion of the virial region in 
each cluster; a drawback is that one requires a model of the potential
to calculate \rfive, and so we also present measurements within a fixed
metric radius to avoid this uncertainty.  We calculate the radius \rfive\ 
and enclosed binding mass \mfive\ two different ways:
(i) using virial scaling relations and (ii) using the isothermal $\beta$ model.
The virial scaling relation is motivated by the simple assumptions that clusters
have recently formed and are approximately virialized and self-similar; in
this case $r_{500}\propto T^{1/2}$ and $M_{500}\propto T^{3/2}$.
Our simulated cluster ensemble exhibits scaling relations
consistent with these assumptions, in agreement with previous 
work (\cite{evrard96,schindler96}).  On the other hand, estimates of
\mfive\ and \rfive\ with the $\beta$ model require only that the ICM 
be approximately in hydrostatic equilibrium.  Rather than using simulated clusters to 
normalize our scaling relation, we normalize using the $\beta$ model results
in the relatively relaxed cluster A\,1795.
\begin{eqnarray}
\label{scaling}
r_{500}=2.37h^{-1}_{50}{\rm\ Mpc}
\left(\left<T_X\right>\over 10\,{\rm KeV}\right)^{1/2}\\
M_{500}=2.00\times10^{15}h^{-1}_{50}M_\odot 
\left(\left<T_X\right>\over 10\,{\rm KeV}\right)^{3/2}\nonumber
\end{eqnarray}
If our $\beta$ model mass for A\,1795 is in error,
we will still be studying comparable regions of each cluster;  those
regions would have an enclosed overdensity different than 500 times the
critical density.

Tables \ref{ICMresults} and \ref{ICMmasses} contain the results.
The columns in Table \ref{ICMresults} correspond to cluster name, ICM central
density $\rho_0$, central electron number density $n_e$, and
the radial location \rave\ of a typical ICM particle located within \rfive. 
Table \ref{ICMmasses} contains the ICM masses \MICM\ and mass fractions 
\fICM.  The columns correspond to cluster name; \MICM\ and \fICM\ calculated 
within a limiting radius $r_{lim}=1h^{-1}_{50}$~Mpc; \rfive, 
\MICM\ and \fICM\ calculated using the virial relation; and \rfive, \MICM\ and
\fICM\ calculated using the isothermal $\beta$ model.
All uncertainties correspond to 90\% confidence limits (see $\S3.5$).

\subsubsection{ICM Masses \MICM\ and Mass Fractions \fICM}

Fig. \ref{ICM500} contains a plot of \MICM\ and \fICM, calculated within
a limiting radius \rfive, versus emission weighted mean ICM temperature \TX.  
The left panel contains results obtained using the isothermal $\beta$ model
to estimate binding masses and \rfive, and the right panel contains 
results using the virial scaling relation (Eqn. \ref{scaling}).
As noted above, measurements within \rfive\ sample the same portion of each
cluster's virial region,
whereas measurements within a fixed metric radius are directly comparable
to results from previous analyses and are more nearly independent of
assumptions about the cluster potential.

Within \rfive\ the best fit relation between \MICM\ and \TX\ is
\begin{equation}
M_{ICM}=(1.49\pm0.09)\times10^{14}\ M_\odot\ T_6^{\,1.98\pm0.18}
\end{equation}
where $T_6$ is \TX\ in units of 6~keV. The variance weighted
RMS scatter in \MICM\ around this relation is 17\%; in calculating 
the variance weighted RMS, each cluster is weighted using its \MICM\ 
measurement uncertainty. 
We find the best fit relation by minimizing $\chi^2$,
the sum of the squared, orthogonal deviations $\nu^2_i$
of each point; $\nu_i$ is the minimum distance between point $i$ and the
fit scaled by one over $\sigma_i$, the point $i$ uncertainty along that vector.
We fit using only those 42 clusters with well determined temperature 
uncertainties, and we determine the 90\% uncertainties on the fit
parameters using 500 bootstrap fits. 

The 17\% scatter about this relation is another indicator of the regularity
of galaxy clusters; the scatter is consistent with constraints derived from 
analysis of scatter in the Size-Temperature relation in an X-ray flux 
limited cluster sample (\cite{mohr97a}) 
and the luminosity-temperature relation in a cooling-flow free 
cluster sample (\cite{arnaud98}).  The slope
of this relation is steep; it is statistically inconsistent with a slope
of ${3\over2}$, the expectation if galaxy clusters are self similar objects.
We also examine the \MICM-\TX\ relation using measurements where \rfive\ is
determined using the virial theorem (see Eqn. \ref{scaling}).  In comparison
to the $\beta$ model analysis, these data exhibit a somewhat smaller
14\% scatter (see Fig \ref{fICMvir}),
but a statistically consistent slope $1.94\pm0.12$.

The best fit \fICM-\TX\ relation within \rfive\ is 
\begin{equation}
f_{ICM}=(0.207\pm0.011)\ T_6^{\,0.34\pm0.22}
\end{equation}
where $T_6$ is \TX\ in units of 6~keV; the variance 
weighted scatter about the relation is 17\%. These data provide
evidence for a modest increase in \fICM\ with \TX.  The slope is inconsistent
with zero at the 99\% confidence level.  We probe for variations in \fICM\
by dividing the sample at $\left<T_X\right>=5$~keV; the weighted mean of 
the 17 low \TX\  (28 high \TX) clusters is 0.167$\pm$0.008 (0.212$\pm$0.006),
corresponding to a statistically significant difference of 0.045$\pm$0.010
(uncertainties are 90\% confidence intervals determined using RMS scatter
of measurements about the mean).  The mean \fICM\ of the low \TX\
sample rises to 0.176$\pm$0.006 if A\,1060 is excluded, bringing the
difference in the mean of the two samples to 0.036$\pm0.001$.
An analysis of \fICM\ where \rfive\ and \mfive\ are calculated using
a virial scaling relation suggests a slightly more rapid increase of \fICM\ with
\TX.  The best fit slope is 0.41$\pm$0.16, and \fICM\ is larger in hot
clusters than in cool ones by 0.052$\pm$0.009.  We caution that the virial
scaling relation assumes self similarity in the cluster population, but the
measurements presented above demonstrate the ICM properties are
inconsistent with self similarity. Therefore, it is a distinct
possibility that the virial scaling relation masses \mfive\ are systematically 
in error.

We have examined the same relations within a fixed radius of
1$h^{-1}_{50}$~Mpc.  The best fit relations between \MICM, \fICM\ and \TX\ are
\begin{eqnarray}
M_{ICM}=(7.45\pm0.49)\times10^{13}\ M_\odot\  T_6^{\,1.23\pm0.17}\\
f_{ICM}=(0.184\pm0.009)\ T_6^{\,0.17\pm0.17}
\end{eqnarray}
where $T_6$ is \TX\ in units of 6~keV.
As expected, the \MICM-\TX\ is shallower than the corresponding 
\rfive\ relation presented above.  The \fICM-\TX\ relation is
flatter and consistent with no increase of \fICM\ with \TX; this is
primarily caused by a reduction in \fICM\ in the higher \TX\ portion
of the sample, behavior that implies an increasing \fICM\ with radius.

\myputfigure{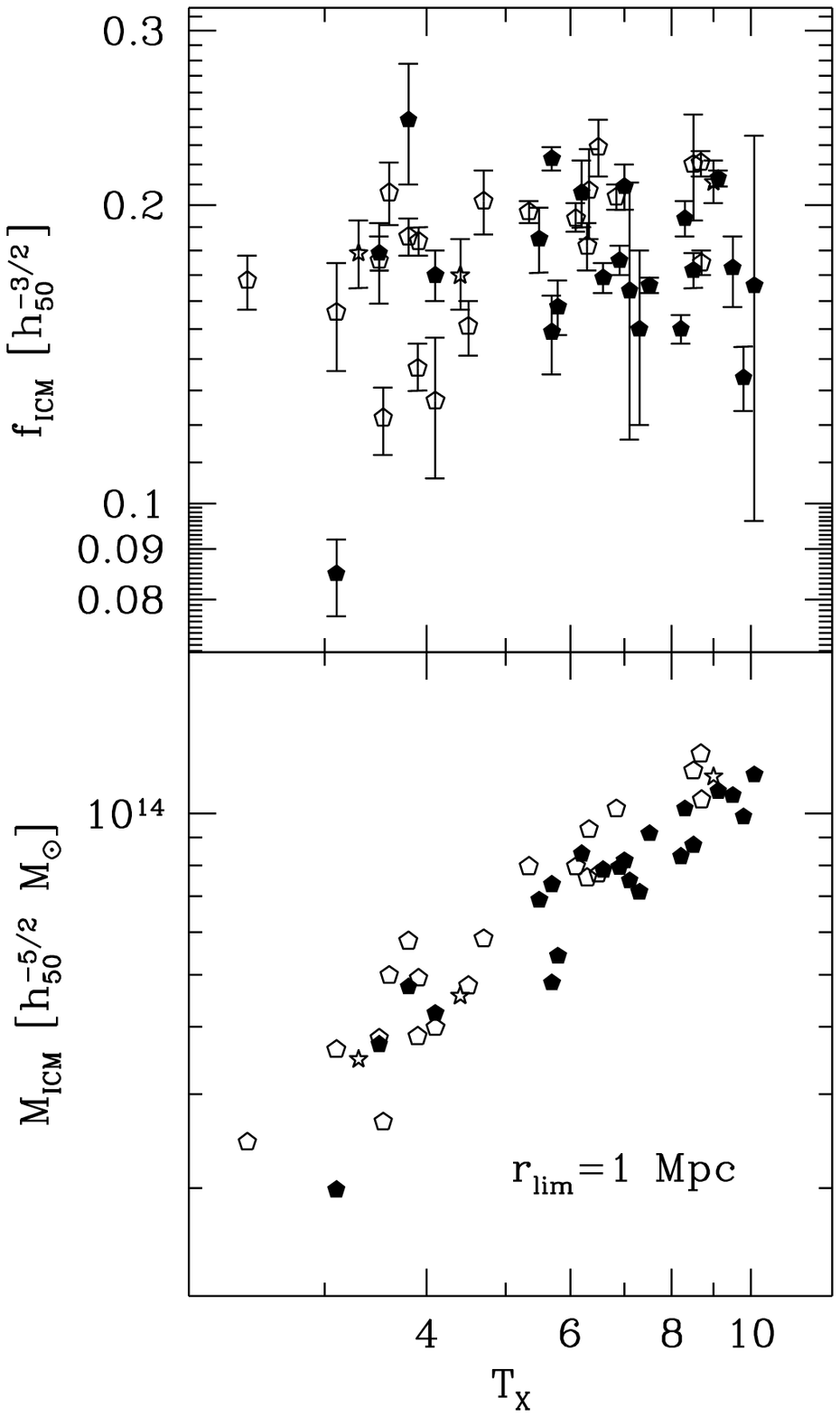}{-3.0}{0.90}{-40}\figcaption{
We plot ICM mass \MICM\ (below) and mass fraction \fICM\ (above) versus \TX;
these quantities are measured within the limiting radius
$r_{lim}=1h_{50}^{-1}$~Mpc.
Clusters with different central cooling times are
represented with different symbols as in Fig. \ref{ICM500}.
}\label{fICMvir}
\medskip

In addition, the data suggest a different
\MICM-\TX\ zeropoint in short and long cooling time clusters; 
the zeropoint for the 20 short cooling time clusters is 
$8.19^{\pm0.81}\times10^{13}$~M$_\odot$
and for the the 22 long cooling time clusters is 
$6.29^{\pm0.43}\times10^{13}$~M$_\odot$.  
This corresponds to $\Delta M_{ICM}=1.90^{\pm0.92}\times10^{13}$~M$_\odot$,
a 3.3$\sigma$ or $\sim$25\% offset.  An offset of similar scale appears in
measurements made within a 0.75$h^{-1}_{50}$~Mpc aperture; 
$\Delta M_{ICM}=1.32^{\pm0.65}\times10^{13}$~M$_\odot$ which is a 3.2$\sigma$
or $\sim$25\% offset.  At the larger radius \rfive\ the offset is
$\Delta M_{ICM}=2.73^{\pm1.91}\times10^{13}$~M$_\odot$ which is a 2.3$\sigma$
or 18\% offset.

This correlation between \MICM\ and central cooling time could be 
because, at a particular mass scale or \TX, clusters with higher gas content
are more likely to experience central cooling instabilities;  in that case
we need to understand what mechanism drives the differences in gas content. 
Our numerical cluster simulations which include gravity and gas dynamics
provide no evidence for large \MICM\ offsets like those observed.
Alternatively, one might suggest that we are seeing the results of
the ICM ``flowing'' into the cluster core;  the scale of the mass offset
requires an average flow rate of $\sim$2000~M$_\odot$/yr over
the age of the universe, which makes this explanation unlikely.
On the other hand, the higher \MICM\ in clusters with shorter central
cooling times may be an artifact; central cooling instabilities may
enable the formation of a multiphase medium and significant
ICM density and temperature variations rather
than the smooth density distribution and isothermal ICM
implicit in the $\beta$ model analysis.  These density variations enhance
the X-ray emissivity, causing an increase in the X-ray luminosity and a bias
in our estimates of the central density $\rho_0$ (see Eqn. \ref{centralrho}).
Quantitatively the offsets in \MICM\ could be explained by a
multiphase medium in short central cooling time clusters if that medium
enhanced the X-ray luminosity by $\sim50$\% on average.

\subsubsection{ICM Radial Distribution}

We also probe for systematic differences in the radial distribution
of the ICM in high and low \TX\ clusters.  The presence of differences 
could be an indication that galactic winds have contributed significant
energy to the ICM (\cite{white91,metzler94,metzler98}).
In past studies, discussion has focused on the 
value $-3\beta$, the asymptotic slope of the ICM; 
we take a new approach which incorporates the entire ICM distribution
within the radius \rfive.
Specifically, we calculate the average radial location \rave\ of an ICM
particle within the same physical region of each cluster, where
\begin{equation}
\left<r\right>= {\int_0^{r_{500}}d^3r\  r\rho(r)\over
\int_0^{r_{500}}d^3r\ \rho(r)}
\end{equation}
Differences in the ICM radial distribution with \TX\ will appear
as trends in \rave\ versus \TX.

Fig. \ref{averadius} (lower panel) contains a plot of
\rave\ versus \TX\ for our cluster sample.  
Symbols code central cooling times as in other figures.
The distribution of \rave\ in this flux limited cluster sample 
provides no indication that the ICM is more extended in low mass clusters.  
These data stand in contrast to previous analyses of large {\it Einstein} IPC 
samples (e.g. \cite{mohr95}) which indicated a tendency for
low \TX\ clusters to have lower $\beta$'s (more extended ICM distributions).

We have analyzed the entire sample with a single $\beta$ model fit, and
we recover a trend for lower \TX\ cluster to have a more extended ICM.
Therefore, the root of the discrepancy between our results and previous
results lies primarily in the treatment of the central emission excesses.
Fig. \ref{doublesinglefit} demonstrates the differences between single and 
double component $\beta$ model fits to the X-ray data on A\,1795, a cluster
with a central emission excess.  We
plot the best fit single component model (dashed line), the double component
model (solid line), the residuals around the fit (single: solid points,
double: hollow points), and we list the best fit parameters and reduced
$\chi^2$.  The behavior in A\,1795 is generic to clusters with central emission
excesses; the best fit single $\beta$ model has a small core radius and
correspondingly low $\beta$ that grossly misrepresents the asymptotic behavior
of the surface brightness profile.  The two extra free parameters in
the double $\beta$ model fit allow the core and asymptotic behavior of the
profile to be fit simultaneously; the gross trends in the residuals
disappear and the goodness of fit parameter $\chi^2$ falls by a factor
of 5.  As discussed in $\S$3.1, adding the second component is similar
in spirit to excluding the core region from the fit; however, results of
fits to core excised clusters depend on the scale of the region excised,
exposing the fit parameters to the subjective whims of the observer.
We choose double $\beta$ model fitting because it is more objective, and
because our data demonstrate that the extra component allows the core
and the asymptotic behavior of the profiles to both be fit
simultaneously.

We also probe for trends in the ICM distribution with \TX\ using
$\beta_{eff}$, a nonparametric measure of the slope of the surface
brightness profile.  Specifically,
\begin{equation}
\beta_{eff} = {1\over3}\left[{\log(I_i/I_o)\over\log(A_o/A_i)} + 
{1\over2}\right];
\end{equation}
operationally, we choose an area $A_o$, find the isophote $I_o$ which encloses
that area, and then find the isophote $I_i$ which encloses the area 
$A_i=0.7A_o$.  This approach yields the effective ICM falloff $-3\beta_{eff}$
over a particular annulus determined by $A_o$.
Fig. \ref{doublesinglefit} (upper panel) contains a 
plot of \betaeff\ measured at $r_{5000}$ versus \TX\ for the same sample; the
data provide no indication that lower \TX\ clusters have a more extended ICM.
Because $\beta_{eff}$ changes with radius in a cluster (small in the core
and steepening with radius) it is important that it be measured in the same
physical region in each cluster (rather than at a particular isophote; see 
\cite{mohr97a}).  The results displayed here are calculated at a radius
$r=0.75h^{-1}_{50}(\left<T_X\right>/10{\rm\ keV})$~Mpc, corresponding 
to $r_{5000}$ and $A_o=\pi r^2_{5000}$,
a radius at which $\beta_{eff}$ can be measured in each cluster image.

\subsection{Comparison to Previous Results}

We compare our \MICM\ and \fICM\ measurements to previously 
published results. Rather than present detailed notes on each cluster,
we focus on two trends apparent from the comparisons.
First, our measurements in the clusters fit with single $\beta$ models are
generally consistent with previously published results.
Second, our measurements on clusters with central emission excesses tend
to produce lower ICM masses \MICM\ than previously published results. 
In individual cases of disagreement in $\beta$ parameters, we invite the 
reader to examine the X-ray surface brightness profiles and contour 
plots (Fig. \ref{profiles}) as a way of gauging the plausibility of 
our fit
(\cite{AFK96,Bardelli96,Breen94,BHB92,BH96,buote96,cirimele97,david95,dellantonio95,fabricant89,HBN93,henriksen96,HJ96,Makino94,NB95,Pislar97,SM97,white95}). 

\myputfigure{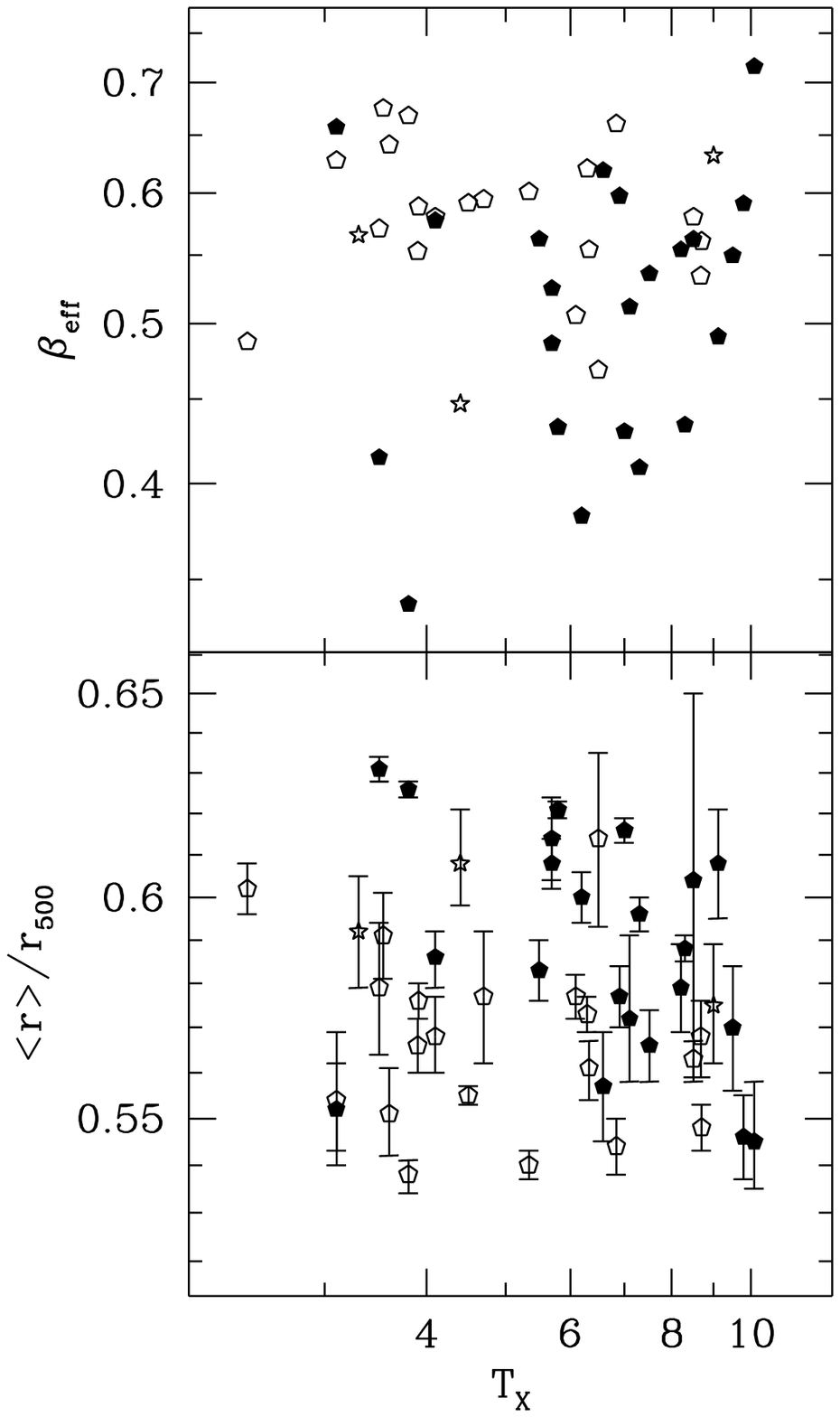}{-3.0}{0.90}{-45}
\figcaption{We plot the typical location of a gas particle
within $r_{500}$ (below) and $\beta_{eff}$, a nonparametric estimate of the
radial falloff of the ICM calculated at $\sim r_{5000}$ (above). 
As before, symbols denote different central cooling times.
Neither dataset provides evidence for a more extended ICM 
in low \TX\ clusters (see discussion in $\S$4.2.2).}
\label{averadius}
\medskip

We have examined the issue of emission excesses in great detail; 
we analyze all clusters with central emission excesses 
using a single and double component $\beta$ 
model.  In the median, the double $\beta$ model analysis provides 
an \MICM\ which is 7\% lower than the single $\beta$ model analysis 
(maximum reduction is 20\% in A\,3526).  This results because the 
emission excess biases the parameters $R_{c}$ and $\beta$ low in the 
single component fits.
The works which reported higher ICM masses generally used a single-beta
model approach and ignored the central emission excesses; in these cases
we were able to demonstrate consistency by showing that we could match
the author's result using a single-beta model ourselves.  A few authors
took the approach of excising the central region before fitting the
cluster profile; these works generally show better agreement with our
double-beta model results.

We also found four papers which do not
follow the trends outlined above. Three of these 
(\cite{durret94,myers97,fukumoto92})
employ very different analyses, making 
it difficult to discern the source of the disagreement.  The recent
deprojection analysis of 207 cooling flows using IPC
data (\cite{white97})
presents ICM masses and mass fractions for many of our clusters and
seems to employ a comparable method, but we found poor agreement
between our results.  There are 18 clusters which we both measure
out to 1$h^{-1}$\,Mpc; out of this set, there are 9 clusters for which
our uncertainty ranges in \MICM\ do not overlap. Strangely, there
is no evidence of a systematic difference in our results; when plotted
against one another the scatter is just much larger than the error bars.
Analysis of the ICM mass fractions \fICM\ is even more discouraging, with
only two clusters in agreement.  Here, however, there is a systematic
difference in that the fractions of WJF are almost all much lower than
our own.  The worst offenders are clusters for which WJF uses older
X-ray temperature determinations (Cygnus A, A\,2029, A\,2142, A\,2319, 
A\,401 and A\,496) or which have no measured velocity dispersions. 
The source of this discrepancy is still unclear, but seems
to be due at least in part to the large uncertainties in the galaxy 
velocity dispersions $\sigma_{gal}$,
which sets the depth of the potential well in their deprojection
technique.  As already mentioned in the introduction, calculating ICM 
masses  and mass fractions does not require assumptions about the detailed 
shape of the cluster potential, and so we make none in our analysis.

To conclude, we reiterate that because the ICM distribution follows
directly from the X-ray surface brightness profile, a model which
provides a better fit to the data will yield a more accurate estimate
of \MICM\ (see $\S3$.  Extensive testing of the double beta model approach
convinces us that the technique produces a significantly better
description of the data, allowing simultaneous fitting of the core 
emission excess and the asymptotic behavior of the gas at the 
limiting radius of the data.

\section{DISCUSSION}

We analyze the X-ray and ICM properties of an
essentially X-ray flux limited sample of 45 galaxy clusters.
The fundamental observations are archival ROSAT PSPC
images and published \TX\ measurements; we describe
our techniques in $\S$3 to facilitate reproducibility.  Combining our data
and an ensemble of numerical cluster simulations, we quantify the
uncertainties in the measured quantities.

In addition to constraining the properties of the ICM in nearby
clusters (see discussion below and $\S$4), we include a set 
(see Fig. \ref{profiles}) of cluster X-ray surface brightness profiles in 
physical units along with contour plots of the cluster X-ray 
emission to help the reader evaluate the goodness of fit, the
importance of substructure and the overall plausibility of the analysis in
each cluster.
While a few clusters in our sample are undergoing major mergers,
making our analysis suspect, it should be remembered that our analysis of
an ensemble of simulated  clusters indicates
that \MICM\ can be estimated with a typical accuracy of $\sim$9\% in
coeval populations with similar morphologies to those observed (i.e. 
even with merging clusters included).

\subsection{ICM Regularity}
The observed scatter around the \MICM-\TX\ and \fICM-\TX\ relation is 17\%;
correcting the scatter for the 9\% observational contribution (derived from
numerical simulations) reduces the scatter to 14\%.
This scatter is consistent with predictions made using the scatter 
around the cluster Size-Temperature relation (\cite{mohr97a}).  
The properties of this X-ray flux limited
cluster sample provide further evidence that galaxy clusters are, on average,
regular objects (see also \cite{mohr97a,arnaud98}).  
Interestingly, ``exceptional'' clusters do exist;
A\,1060 has an \MICM\ and \fICM\ which are low relative to clusters with 
similar \TX\ (see also \cite{lowen96}).

\subsection{Self-similarity and Evidence for Galaxy Feedback}
Measurements of \MICM\ at \rfive\ provide evidence that the ICM 
content of galaxy clusters
is not self-similar.  Specifically, the slope of the \MICM-\TX\ relation
at \rfive\ is 1.98$\pm$0.18 (90\% confidence unless otherwise noted),
different from the self-similarity expectation
of ${3\over2}$ at 4.3$\sigma$.  We also probe for trends in \fICM\ with \TX.
The best fit \fICM-\TX\ relation has a slope of 0.34$\pm$0.22, and is
inconsistent with no variation at 2.5$\sigma$.  If the steep slope of
the \MICM-\TX\ relation were entirely caused by an increase of \fICM\ with
\TX, the required \fICM-\TX\ relation slope would be 0.48, well within the 
90\% confidence interval of our measurements; therefore, it is 
possible that the steep \MICM-\TX\ slope is due solely to an increase 
of \fICM\ with \TX.  We also examine trends in \fICM\ by comparing
clusters with \TX\ above and below 5~keV.  We find that the 28 clusters with 
$\left<T_X\right>>5$~keV have \fICM=0.212$\pm$0.006, 0.045$\pm$0.010 higher
than the mean of the 17 low \TX\ clusters.  Thus, these data indicate that
\fICM\ increases with \TX, and that this increase may be responsible for
the steepness of the \MICM-\TX\ relation.  

This departure from self similarity in the ICM properties of clusters could
be a reflection of a lack of self similarity in the underlying dark matter
distribution; the ``universal'' NFW density profile predicts
significant departures from self similarity in the cluster population driven
by earlier formation, on average, of low mass clusters when compared to high
mass clusters (\cite{navarro97}).  On the other hand, the ICM is sensitive to
a wider range of physics than the dark matter, and this sensitivity might,
in part, be responsible for the observed departures from self similarity.
Energy injection into the intergalactic medium by supernovae driven galactic
winds is one possible mechanism; a trend of increasing \fICM\ with \TX\ is
consistent with some models of galaxy feedback (\cite{metzler98,tozzi98}).
Another possibility is that the efficiency of galaxy formation varies 
with cluster mass (\cite{david90,arnaud92}).

Interestingly, our extensive dataset provides no evidence that the
ICM is more extended in low \TX\ clusters.  Specifically, there are no 
apparent trends in the typical radial location \rave\ of an ICM
particle within \rfive,
a non-parametric estimate of the radial fall-off of the ICM density
$\beta_{eff}$ at $r_{5000}$, or $\beta$ measured fitting to the whole
surface brightness profile.  Previous analyses of large samples have
relied primarily on single $\beta$ model fitting,
which we show (see Fig. \ref{doublesinglefit} and $\S$4.2)
returns highly biased estimates of $\beta$ and $r_c$ in clusters with
large, central emission excesses.  This bias, combined with the greater 
prevalence of
low \TX\ clusters with central emission excesses, leads to an artificial
trend in $\beta$ or \rave\ with \TX.  We recover this trend when
we analyze our entire sample using solely single $\beta$ model fits.
Models of galaxy feedback generally produce a more extended ICM in low
mass clusters (\cite{metzler98}). While our data do not rule out a
weak trend in ICM extent, they provide no suggestion that one exists 
above 2.4\,keV.

\subsection{Constraints on $\Omega_M$}
Under the fair sample hypothesis, baryon fraction measurements within cluster
virial regions can be used to derive an unbiased estimate
of the universal baryon fraction.
By combining measurements of \fICM\ in clusters 
(a lower limit on the cluster baryon fraction) with
nucleosynthesis constraints on the baryon to photon ratio, we
place an upper limit on the cosmological density parameter for
clustered matter $\Omega_M$
(e.g. \cite{white93}).  

Because more massive clusters are less affected by processes like galaxy 
feedback which might enhance ICM depletion within the cluster virial region,
we use the mean \fICM\ calculated in the
$\left<T_X\right>>5$~keV subsample of 27 clusters: \fICM=0.212$h^{-3/2}_{50}$
(see $\S$4.2.1).  We apply two corrections to this mean \fICM: 
(1) a correction for ICM depletion within \rfive\ (driven by
shocks during cluster formation), and
(2) a correction for ICM clumping which causes an overestimate of \fICM.
We also use the numerical simulations to estimate ICM depletion within
\rfive; our ensemble of 48 simulations indicates that, on average,
\fICM\ within \rfive\ is 12\% lower than the cosmic value.
As discussed in $\S3.5$ and $\S4.2.1$, ICM density variations or clumping
enhance the X-ray luminosity and bias \fICM\ estimates high.  Analysis of
our numerical simulations indicates that, on average, estimates of
\MICM\ and \fICM\ are too high by 12\%.  There is no guarantee that the
structure responsible for this overestimate in our simulations mimics
the structure present in real clusters.  Hot galactic
winds, magnetic fields, and radiative cooling instabilities could well
enhance density structure; in fact, larger variations in the ICM 
density distribution, and therefore larger systematic errors in \MICM, 
are possible if the ICM has a strongly multi-phase structure 
(\cite{gunn96}, Nagai, Sulkanen \& Evrard in prep).
Comparisons of cluster X-ray emission and the Sunyaev-Zel'dovich effect 
decrement in the cosmic microwave background should
further constrain the density and temperature variations in the ICM.

Correcting the mean \fICM\ for clumping and depletion
(these corrections cancel), we
get our best estimate of the cosmic ICM mass fraction:
$f_{ICM}=(0.212\pm0.006)h^{-3/2}_{50}$, where the uncertainty represents
the 90\% confidence range from observational uncertainties ($\S3.5$).
Constraints on the clustered mass density parameter $\Omega_M$ are then
\begin{equation}
\Omega_M<{\Omega_b\over \left<f_{ICM}\right>}=
(0.36\pm0.01)\Theta_{sys}h_{50}^{-1/2}
\end{equation}
where we have used $\Omega_b=(0.077\pm0.001)h_{50}^{-2}$ (\cite{burles98}).
The $\sim3$\% uncertainty reflects the observational uncertainty of the mean
\fICM; $\Theta_{sys}$ is a scale factor that represents
systematic effects which we haven't addressed.
Our analysis of an ensemble of 48 numerical cluster simulations suggests
that after correcting for depletion and gas clumping,
$\Theta_{sys}=1$.  Deviations of $\Theta_{sys}$ from 1
would reflect systematics which appear in real clusters but not 
in simulated clusters.  Some possibilities are (1) very steep
temperature profiles at \rfive\ and (2) a very strong multiphase medium
extending throughout the cluster.  Recent comparisons of lensing mass estimates
and isothermal $\beta$ model binding mass estimates
also suggest $\Theta_{sys}\sim1$;
the mean ratio of isothermal $\beta$ model to lensing binding 
masses is 1.04$\pm$0.07 for the CNOC cluster sample (\cite{lewis98}).  
Additionally, cluster ICM masses \MICM\ have an associated systematic
uncertainty due to uncertainties in the absolute flux calibration of
the ROSAT PSPC; we estimate that the PSPC absolute
flux calibration is good to $\sim15$\%,
implying that ICM masses are uncertain by approximately 7.5\%. 
Combining these estimates
of systematic uncertainties yields $\Theta_{sys}=1\pm0.10$.
Future binding mass analyses of even larger cluster samples and
observations from the well calibrated AXAF satellite (\cite{elsner94})
will further constrain $\Theta_{sys}$.

\acknowledgements

We acknowledge helpful discussions with Michael Turner.
This research makes use of the NASA/GSFC HEASARC Online Service,
and benefits significantly from the efforts of the
PROS software development team at the Smithsonian Astrophysical
Observatory.
This work is supported by NASA grants NAG5-3401 and NAGW-2367 and is
supported through AXAF Fellowship grant PF8-1003, 
awarded through the AXAF Science Center.  The AXAF Science Center
is operated by the Smithsonian Astrophysical Observatory for
NASA under contract NAS8-39073.

\begin{deluxetable}{llrr}
\tablewidth{0pt}
\tablecaption{Cluster Image Information and \TX}
\tablehead{
\colhead{Cluster}	&
\colhead{Sequence \#}	&
\colhead{$t_{exp}$}	&
\colhead{\TX}}
\scriptsize
\startdata
A\,85        & wp800174        & 14778 & 6.10$^{+0.12}_{-0.12}$ \nl
\multicolumn{4}{l}{\hskip0.5cm rp800250} \nl  
A\,119       & rp800251        & 14281 & 5.80$^{+0.36}_{-0.36}$ \nl
A\,262       & rp800254        &  8063 & 2.41$^{+0.03}_{-0.03}$ \nl
A\,401       & wp800182        & 13408 & 8.30$^{+0.31}_{-0.31}$ \nl
\multicolumn{4}{l}{\hskip0.5cm wp800235} \nl  
A\,426       & rp800186        &  4435 & 6.33$^{+0.21}_{-0.18}$ \nl
A\,478       & wp800193        & 21345 & 6.84$^{+0.13}_{-0.13}$ \nl
A\,496       & rp800024        &  8117 & 3.91$^{+0.04}_{-0.04}$ \nl
A\,644       & wp800379n00     &  9535 & 6.59$^{+0.10}_{-0.10}$ \nl
A\,754       & rp800232n00     & 16125 & 8.50$^{+0.30}_{-0.30}$ \nl
\multicolumn{4}{l}{\hskip0.5cm wp800160, wp800550} \nl  
A\,780       & rp800318n00     & 17255 & 3.80$^{+0.12}_{-0.12}$ \nl
A\,1060      & wp800200        & 14563 & 3.10$^{+0.15}_{-0.15}$ \nl
A\,1367      & rp800153n00     & 17705 & 3.50$^{+0.11}_{-0.11}$ \nl
A\,1651      & wp800353        &  7155 & 6.30$^{+0.30}_{-0.30}$ \nl
A\,1656      & rp800005        & 19401 & 8.21$^{+0.16}_{-0.16}$ \nl
A\,1689      & rp800248        & 13300 & 10.10$^{+5.40}_{-2.80}$ \nl
A\,1795      & rp800055        & 58809 & 5.34$^{+0.07}_{-0.07}$ \nl
\multicolumn{4}{l}{\hskip0.5cm rp800105} \nl  
A\,2029      & rp800249        & 14767 & 8.70$^{+0.18}_{-0.18}$ \nl
\multicolumn{4}{l}{\hskip0.5cm wp800161} \nl  
A\,2052      & rp800275        &  8331 & 3.10$^{+0.20}_{-0.20}$ \nl
\multicolumn{4}{l}{\hskip0.5cm rp800275a01}\nl  
A\,2063      & wp800184        &  9613 & 4.10$^{+0.60}_{-0.60}$ \nl
A\,2142      & rp800233        & 23154 & 8.68$^{+0.12}_{-0.12}$ \nl
\multicolumn{4}{l}{\hskip0.5cm wp150084, wp800096, wp800551n00}\nl  
A\,2199      & wp150083        & 47162 & 4.50$^{+0.20}_{-0.10}$ \nl
\multicolumn{4}{l}{\hskip0.5cm wp800644n00} \nl  
A\,2204      & rp800281        &  5075 & 9.00 \nl
A\,2244      & rp800265n00     &  2863 & 7.10$^{+2.40}_{-1.50}$ \nl
A\,2255      & rp800512n00     & 12457 & 7.30$^{+1.70}_{-1.10}$ \nl
A\,2256      & wp100110        & 17105 & 7.51$^{+0.11}_{-0.11}$ \nl
A\,2319      & wp800073        &  3932 & 9.12$^{+0.09}_{-0.09}$ \nl
\multicolumn{4}{l}{\hskip0.5cm wp800073a01} \nl  
A\,2597      & rp800112n00     &  6894 & 3.60$^{+0.12}_{-0.12}$ \nl
A\,3112      & rp800302n00     &  7240 & 4.70$^{+0.24}_{-0.24}$ \nl
A\,3158      & rp800310        &  2921 & 5.50$^{+0.30}_{-0.40}$ \nl
A\,3266      & rp800211n00     & 19810 & 6.20$^{+0.50}_{-0.40}$ \nl
\multicolumn{4}{l}{\hskip0.5cm wp800552n00} \nl  
A\,3391      & wp800080        &  5529 & 5.70$^{+0.42}_{-0.42}$ \nl
A\,3526      & rp800607a01     &  9480 & 3.54$^{+0.08}_{-0.08}$ \nl
\multicolumn{4}{l}{\hskip0.5cm wp800192} \nl  
A\,3532      & wp701155n00     &  7742 & 4.40 \nl
A\,3558      & rp800076n00     & 27788 & 5.70$^{+0.12}_{-0.12}$ \nl
A\,3562      & rp800237n00     & 18497 & 3.80$^{+0.50}_{-0.50}$ \nl
A\,3571      & rp800287        &  5465 & 6.90$^{+0.18}_{-0.18}$ \nl
A\,3667      & rp800234n00     & 11328 & 7.00$^{+0.36}_{-0.36}$ \nl
A\,4038      & wp800354n00     &  3207 & 3.30 \nl
A\,4059      & wp800175        &  5221 & 4.10$^{+0.18}_{-0.18}$ \nl
0745\,-19        & rp800623n00     &  9050 & 8.50$^{+1.20}_{-0.80}$ \nl
AWM\,7            & wp800168        & 12603 & 3.90$^{+0.12}_{-0.12}$ \nl
Cygnus\,A        & wp800622n00     &  8733 & 6.50$^{+0.36}_{-0.36}$ \nl
MKW\,3S          & rp800128        &  9228 & 3.50$^{+0.12}_{-0.12}$ \nl
Ophiucus        & rp800279n00     &  3713 & 9.80$^{+0.61}_{-0.61}$ \nl
Tria Aust       & rp800280n00     &  6640 & 9.50$^{+0.42}_{-0.42}$ \nl
\enddata
\label{basicdata}
\end{deluxetable}

\begin{deluxetable}{lcccccrrr}
\tablewidth{0pt}
\tablecaption{Parameters for X--ray Surface Brightness Profiles}
\tablehead{
		&
\colhead{$I_1$\tablenotemark{a}}	&
\colhead{$R_1$}	&
\colhead{$I_2$\tablenotemark{a}}	&
\colhead{$R_2$}	&
\cr
\colhead{Cluster} &
\colhead{ergs/s/cm$^2$/\sq\arcmin}	&
\colhead{$h_{50}^{-1}$Mpc}        	&
\colhead{ergs/s/cm$^2$/\sq\arcmin}	&
\colhead{$h_{50}^{-1}$Mpc}        	&
\colhead{$\beta$} &
\colhead{$\chi^2_\nu$} &
\colhead{\#}    &
\colhead{$\theta_{m}$[\arcmin]}}
\scriptsize
\startdata
A\,85      &  3.92$^{0.54}_{0.60}$e-13 &  0.317$^{0.054}_{0.047}$ &  4.92$^{0.79}_{0.63}$e-12 &  0.063$^{0.011}_{0.009}$ &  0.662$^{0.029}_{0.024}$ &   3.39 &  81 & 19.9 \nl
A\,119     &  1.18$^{0.04}_{0.04}$e-13 &  0.483$^{0.028}_{0.028}$ &                                     &                          &  0.662$^{0.023}_{0.022}$ &   1.21 &  97 & 23.7 \nl
A\,262     &  1.59$^{0.48}_{0.46}$e-13 &  0.135$^{0.035}_{0.034}$ &  2.87$^{0.89}_{0.85}$e-12 &  0.014$^{0.004}_{0.004}$ &  0.556$^{0.027}_{0.025}$ &   1.75 &  97 & 23.7 \nl
A\,401     &  1.01$^{0.08}_{0.05}$e-12 &  0.237$^{0.016}_{0.019}$ &                                     &                          &  0.606$^{0.015}_{0.016}$ &   0.66 &  77 & 18.7 \nl
A\,426     &  5.02$^{0.50}_{0.70}$e-13 &  0.402$^{0.035}_{0.034}$ &  2.14$^{0.19}_{0.19}$e-11 &  0.057$^{0.005}_{0.005}$ &  0.748$^{0.034}_{0.027}$ &   1.95 & 142 & 36.6 \nl
A\,478     &  9.38$^{1.76}_{1.62}$e-13 &  0.269$^{0.045}_{0.045}$ &  9.83$^{1.65}_{1.62}$e-12 &  0.072$^{0.011}_{0.011}$ &  0.713$^{0.030}_{0.033}$ &   5.10 &  56 & 13.7 \nl
A\,496     &  3.78$^{0.44}_{0.33}$e-13 &  0.219$^{0.017}_{0.019}$ &  7.31$^{0.72}_{0.85}$e-12 &  0.033$^{0.003}_{0.003}$ &  0.650$^{0.021}_{0.019}$ &   1.70 &  83 & 20.4 \nl
A\,644     &  1.70$^{0.27}_{0.26}$e-12 &  0.176$^{0.035}_{0.034}$ &                                     &                          &  0.660$^{0.048}_{0.048}$ &   1.52 &  53 & 12.7 \nl
A\,754     &  3.89$^{0.85}_{0.85}$e-13 &  0.367$^{0.324}_{0.309}$ &                                     &                          &  0.614$^{0.361}_{0.360}$ &   9.33 &  92 & 22.4 \nl
A\,780     &  7.03$^{1.50}_{0.42}$e-13 &  0.237$^{0.013}_{0.024}$ &  1.11$^{0.05}_{0.18}$e-11 &  0.047$^{0.002}_{0.008}$ &  0.766$^{0.021}_{0.025}$ &   2.54 &  57 & 15.0 \nl
A\,1060    &  1.98$^{0.15}_{0.16}$e-13 &  0.164$^{0.017}_{0.015}$ &  3.70$^{0.40}_{0.30}$e-13 &  0.043$^{0.005}_{0.004}$ &  0.703$^{0.044}_{0.036}$ &   1.13 & 114 & 27.9 \nl
A\,1367    &  8.41$^{0.41}_{0.40}$e-14 &  0.360$^{0.046}_{0.043}$ &                                     &                          &  0.607$^{0.044}_{0.042}$ &   1.64 & 132 & 32.4 \nl
A\,1651    &  1.45$^{0.09}_{0.07}$e-12 &  0.160$^{0.009}_{0.009}$ &                                     &                          &  0.616$^{0.012}_{0.013}$ &   1.62 &  43 & 10.5 \nl
A\,1656    &  4.65$^{0.14}_{0.14}$e-13 &  0.386$^{0.026}_{0.026}$ &                                     &                          &  0.705$^{0.046}_{0.046}$ &   1.38 &  99 & 24.9 \nl
A\,1689    &  5.15$^{0.64}_{0.77}$e-12 &  0.131$^{0.022}_{0.014}$ &                                     &                          &  0.648$^{0.035}_{0.024}$ &   6.24 &  25 &  6.2 \nl
A\,1795    &  4.96$^{1.03}_{0.43}$e-13 &  0.344$^{0.024}_{0.035}$ &  7.48$^{0.49}_{0.84}$e-12 &  0.083$^{0.005}_{0.009}$ &  0.790$^{0.031}_{0.032}$ &   7.40 &  74 & 17.9 \nl
A\,2029    &  5.63$^{1.37}_{0.48}$e-13 &  0.334$^{0.022}_{0.036}$ &  1.06$^{0.06}_{0.11}$e-11 &  0.079$^{0.004}_{0.009}$ &  0.705$^{0.030}_{0.028}$ &   2.96 &  56 & 14.0 \nl
A\,2052    &  3.35$^{0.95}_{0.60}$e-13 &  0.203$^{0.045}_{0.042}$ &  5.38$^{1.19}_{1.29}$e-12 &  0.039$^{0.009}_{0.009}$ &  0.712$^{0.081}_{0.069}$ &   1.71 &  52 & 13.0 \nl
A\,2063    &  2.39$^{0.70}_{0.33}$e-13 &  0.255$^{0.045}_{0.044}$ &  8.94$^{1.52}_{3.00}$e-13 &  0.059$^{0.012}_{0.020}$ &  0.706$^{0.051}_{0.047}$ &   0.83 &  72 & 17.2 \nl
A\,2142    &  2.71$^{1.03}_{0.95}$e-13 &  0.635$^{0.213}_{0.198}$ &  3.66$^{1.27}_{1.19}$e-12 &  0.164$^{0.054}_{0.051}$ &  0.787$^{0.082}_{0.093}$ &   3.81 &  76 & 18.7 \nl
A\,2199    &  7.84$^{0.74}_{0.58}$e-13 &  0.162$^{0.010}_{0.008}$ &  4.34$^{0.22}_{0.29}$e-12 &  0.041$^{0.002}_{0.002}$ &  0.663$^{0.012}_{0.008}$ &   4.54 & 116 & 28.4 \nl
A\,2204    &  3.14$^{1.74}_{1.48}$e-13 &  0.260$^{0.106}_{0.098}$ &  4.28$^{2.28}_{2.00}$e-11 &  0.035$^{0.015}_{0.012}$ &  0.585$^{0.061}_{0.045}$ &   3.34 &  30 &  6.7 \nl
A\,2244    &  2.36$^{0.75}_{0.70}$e-12 &  0.117$^{0.047}_{0.031}$ &                                     &                          &  0.594$^{0.061}_{0.045}$ &   1.15 &  34 &  8.7 \nl
A\,2255    &  1.68$^{0.08}_{0.08}$e-13 &  0.584$^{0.046}_{0.039}$ &                                     &                          &  0.792$^{0.050}_{0.043}$ &   0.98 &  60 & 14.5 \nl
A\,2256    &  4.41$^{0.10}_{0.10}$e-13 &  0.486$^{0.042}_{0.041}$ &                                     &                          &  0.828$^{0.062}_{0.061}$ &   1.17 &  93 & 22.7 \nl
A\,2319    &  1.14$^{0.21}_{0.21}$e-12 &  0.213$^{0.070}_{0.069}$ &                                     &                          &  0.536$^{0.061}_{0.060}$ &   2.92 &  85 & 20.7 \nl
A\,2597    &  1.51$^{0.26}_{0.27}$e-11 &  0.044$^{0.007}_{0.006}$ &                                     &                          &  0.612$^{0.023}_{0.021}$ &   1.24 &  28 &  7.7 \nl
A\,3112    &  8.55$^{4.89}_{4.67}$e-12 &  0.049$^{0.022}_{0.021}$ &                                     &                          &  0.562$^{0.040}_{0.040}$ &   3.17 &  48 & 12.0 \nl
A\,3158    &  5.42$^{0.54}_{0.51}$e-13 &  0.262$^{0.034}_{0.031}$ &                                     &                          &  0.657$^{0.041}_{0.036}$ &   1.02 &  54 & 13.7 \nl
A\,3266    &  3.09$^{0.28}_{0.27}$e-13 &  0.495$^{0.040}_{0.038}$ &                                     &                          &  0.744$^{0.039}_{0.037}$ &  11.02 &  92 & 22.4 \nl
A\,3391    &  2.30$^{0.31}_{0.26}$e-13 &  0.216$^{0.045}_{0.038}$ &                                     &                          &  0.541$^{0.048}_{0.044}$ &   0.80 &  55 & 13.2 \nl
A\,3526    &  2.01$^{0.72}_{0.67}$e-13 &  0.138$^{0.045}_{0.040}$ &  8.52$^{3.72}_{3.08}$e-12 &  0.012$^{0.005}_{0.004}$ &  0.569$^{0.035}_{0.036}$ &   2.90 & 138 & 33.6 \nl
A\,3532    &  2.04$^{0.37}_{0.29}$e-13 &  0.246$^{0.073}_{0.057}$ &                                     &                          &  0.589$^{0.086}_{0.062}$ &   0.95 &  50 & 12.0 \nl
A\,3558    &  6.89$^{0.54}_{0.51}$e-13 &  0.194$^{0.028}_{0.027}$ &                                     &                          &  0.548$^{0.029}_{0.029}$ &  12.40 &  82 & 19.9 \nl
A\,3562    &  5.62$^{0.42}_{0.34}$e-13 &  0.097$^{0.006}_{0.007}$ &                                     &                          &  0.470$^{0.007}_{0.007}$ &   1.38 &  79 & 19.2 \nl
A\,3571    &  1.46$^{0.10}_{0.10}$e-12 &  0.173$^{0.016}_{0.015}$ &                                     &                          &  0.610$^{0.024}_{0.024}$ &   1.98 &  84 & 21.2 \nl
A\,3667    &  4.84$^{0.38}_{0.37}$e-13 &  0.258$^{0.026}_{0.025}$ &                                     &                          &  0.541$^{0.016}_{0.016}$ &   1.73 &  98 & 23.9 \nl
A\,4038    &  1.84$^{0.32}_{0.29}$e-12 &  0.056$^{0.012}_{0.011}$ &                                     &                          &  0.537$^{0.033}_{0.033}$ &   1.35 &  63 & 15.5 \nl
A\,4059    &  1.61$^{0.22}_{0.20}$e-12 &  0.078$^{0.011}_{0.011}$ &                                     &                          &  0.558$^{0.019}_{0.020}$ &   1.51 &  51 & 12.5 \nl
0745\,-19    &  3.03$^{0.62}_{0.53}$e-11 &  0.056$^{0.006}_{0.006}$ &                                     &                          &  0.586$^{0.008}_{0.009}$ &   2.89 &  36 &  9.5 \nl
AWM\,7       &  4.10$^{0.39}_{0.42}$e-13 &  0.195$^{0.013}_{0.011}$ &  1.36$^{0.22}_{0.14}$e-12 &  0.032$^{0.005}_{0.002}$ &  0.678$^{0.030}_{0.029}$ &   2.58 & 130 & 31.9 \nl
Cygnus\,A       &  3.97$^{2.00}_{2.50}$e-11 &  0.015$^{0.049}_{0.015}$ &                                     &                          &  0.472$^{0.057}_{0.057}$ &   2.42 &  63 & 16.2 \nl
MKW\,3S      &  2.90$^{0.73}_{0.69}$e-12 &  0.056$^{0.015}_{0.015}$ &                                     &                          &  0.562$^{0.038}_{0.038}$ &   2.60 &  44 & 10.7 \nl
Ophiuchus       &  1.27$^{0.29}_{0.15}$e-12 &  0.266$^{0.028}_{0.026}$ &  1.97$^{0.21}_{0.61}$e-12 &  0.079$^{0.008}_{0.025}$ &  0.705$^{0.036}_{0.032}$ &   0.81 & 112 & 27.4 \nl
Tria Aust       &  3.06$^{0.88}_{0.41}$e-13 &  0.621$^{0.076}_{0.088}$ &  8.07$^{0.83}_{1.44}$e-13 &  0.212$^{0.025}_{0.038}$ &  0.816$^{0.062}_{0.060}$ &   0.71 &  87 & 21.2 \nl
\enddata
\tablenotetext{a}{Rest frame within 0.5-2.0~keV band}
\label{SBresults}
\end{deluxetable}

\begin{deluxetable}{lccc}
\tablewidth{0pt}
\tablecaption{ICM Density and Distribution}
\tablehead{
			&
\colhead{$n_e$}		&
\colhead{$\rho_0$}	&
\cr
\colhead{Cluster} 			&
\colhead{$h_{50}^{1/2}$ g cm$^{-3}$}	&
\colhead{$h_{50}^{1/2}$ cm$^{-3}$}	&
\colhead{$\left<r\right>/r_{500}$}
}
\scriptsize
\startdata
A\,85      &   2.62$^{0.17}_{0.22}$e-2 &  5.10$^{0.32}_{0.44}$e-26 &  0.577$^{0.005}_{0.005}$ \nl
A\,119     &   1.40$^{0.04}_{0.04}$e-3 &  2.72$^{0.09}_{0.08}$e-27 &  0.621$^{0.002}_{0.002}$ \nl
A\,262     &   3.74$^{0.28}_{0.19}$e-2 &  7.29$^{0.55}_{0.38}$e-26 &  0.602$^{0.006}_{0.006}$ \nl
A\,401     &   5.87$^{0.43}_{0.27}$e-3 &  1.14$^{0.08}_{0.05}$e-26 &  0.588$^{0.003}_{0.003}$ \nl
A\,426     &   5.50$^{0.27}_{0.45}$e-2 &  1.07$^{0.05}_{0.09}$e-25 &  0.561$^{0.007}_{0.006}$ \nl
A\,478     &   3.81$^{0.33}_{0.15}$e-2 &  7.42$^{0.64}_{0.30}$e-26 &  0.544$^{0.006}_{0.006}$ \nl
A\,496     &   4.18$^{0.33}_{0.14}$e-2 &  8.15$^{0.64}_{0.26}$e-26 &  0.576$^{0.004}_{0.004}$ \nl
A\,644     &   9.17$^{0.40}_{0.39}$e-3 &  1.79$^{0.08}_{0.08}$e-26 &  0.557$^{0.012}_{0.012}$ \nl
A\,754     &   2.92$^{0.08}_{0.08}$e-3 &  5.68$^{0.15}_{0.16}$e-27 &  0.604$^{0.046}_{0.046}$ \nl
A\,780     &   4.77$^{0.68}_{0.20}$e-2 &  9.30$^{1.32}_{0.40}$e-26 &  0.538$^{0.004}_{0.003}$ \nl
A\,1060    &   8.50$^{0.51}_{0.54}$e-3 &  1.66$^{0.10}_{0.11}$e-26 &  0.552$^{0.009}_{0.010}$ \nl
A\,1367    &   1.25$^{0.04}_{0.03}$e-3 &  2.44$^{0.08}_{0.07}$e-27 &  0.631$^{0.003}_{0.003}$ \nl
A\,1651    &   8.80$^{0.40}_{0.39}$e-3 &  1.71$^{0.08}_{0.08}$e-26 &  0.573$^{0.004}_{0.004}$ \nl
A\,1656    &   3.12$^{0.04}_{0.05}$e-3 &  6.07$^{0.07}_{0.09}$e-27 &  0.579$^{0.010}_{0.010}$ \nl
A\,1689    &   2.17$^{0.23}_{0.24}$e-2 &  4.22$^{0.45}_{0.47}$e-26 &  0.545$^{0.010}_{0.013}$ \nl
A\,1795    &   2.99$^{0.46}_{0.15}$e-2 &  5.83$^{0.89}_{0.29}$e-26 &  0.540$^{0.003}_{0.003}$ \nl
A\,2029    &   3.67$^{0.59}_{0.18}$e-2 &  7.14$^{1.14}_{0.36}$e-26 &  0.548$^{0.005}_{0.005}$ \nl
A\,2052    &   3.42$^{0.67}_{0.28}$e-2 &  6.67$^{1.31}_{0.55}$e-26 &  0.554$^{0.014}_{0.015}$ \nl
A\,2063    &   1.16$^{0.24}_{0.09}$e-2 &  2.25$^{0.46}_{0.17}$e-26 &  0.568$^{0.008}_{0.009}$ \nl
A\,2142    &   1.58$^{0.17}_{0.24}$e-2 &  3.08$^{0.34}_{0.46}$e-26 &  0.568$^{0.009}_{0.008}$ \nl
A\,2199    &   2.92$^{0.18}_{0.15}$e-2 &  5.68$^{0.36}_{0.30}$e-26 &  0.555$^{0.002}_{0.002}$ \nl
A\,2204    &   1.17$^{0.17}_{0.13}$e-1 &  2.27$^{0.34}_{0.26}$e-25 &  0.575$^{0.013}_{0.014}$ \nl
A\,2244    &   1.32$^{0.19}_{0.29}$e-2 &  2.58$^{0.37}_{0.57}$e-26 &  0.572$^{0.014}_{0.019}$ \nl
A\,2255    &   1.74$^{0.07}_{0.06}$e-3 &  3.38$^{0.14}_{0.13}$e-27 &  0.596$^{0.004}_{0.004}$ \nl
A\,2256    &   3.02$^{0.05}_{0.05}$e-3 &  5.89$^{0.10}_{0.09}$e-27 &  0.566$^{0.008}_{0.008}$ \nl
A\,2319    &   6.21$^{0.23}_{0.36}$e-3 &  1.21$^{0.05}_{0.07}$e-26 &  0.608$^{0.013}_{0.013}$ \nl
A\,2597    &   5.19$^{0.73}_{0.75}$e-2 &  1.01$^{0.14}_{0.15}$e-25 &  0.551$^{0.009}_{0.010}$ \nl
A\,3112    &   3.61$^{0.46}_{0.30}$e-2 &  7.04$^{0.89}_{0.58}$e-26 &  0.577$^{0.015}_{0.015}$ \nl
A\,3158    &   4.15$^{0.28}_{0.26}$e-3 &  8.08$^{0.54}_{0.51}$e-27 &  0.583$^{0.007}_{0.007}$ \nl
A\,3266    &   2.38$^{0.07}_{0.07}$e-3 &  4.63$^{0.14}_{0.14}$e-27 &  0.600$^{0.006}_{0.006}$ \nl
A\,3391    &   2.73$^{0.30}_{0.28}$e-3 &  5.31$^{0.59}_{0.54}$e-27 &  0.614$^{0.010}_{0.010}$ \nl
A\,3526    &   6.81$^{0.78}_{0.43}$e-2 &  1.33$^{0.15}_{0.08}$e-25 &  0.591$^{0.010}_{0.010}$ \nl
A\,3532    &   2.49$^{0.34}_{0.27}$e-3 &  4.86$^{0.66}_{0.53}$e-27 &  0.608$^{0.010}_{0.013}$ \nl
A\,3558    &   4.95$^{0.18}_{0.15}$e-3 &  9.65$^{0.34}_{0.29}$e-27 &  0.608$^{0.006}_{0.006}$ \nl
A\,3562    &   5.88$^{0.39}_{0.26}$e-3 &  1.15$^{0.08}_{0.05}$e-26 &  0.626$^{0.002}_{0.002}$ \nl
A\,3571    &   7.92$^{0.23}_{0.29}$e-3 &  1.54$^{0.05}_{0.06}$e-26 &  0.577$^{0.007}_{0.007}$ \nl
A\,3667    &   3.64$^{0.13}_{0.13}$e-3 &  7.09$^{0.25}_{0.25}$e-27 &  0.616$^{0.003}_{0.003}$ \nl
A\,4038    &   1.43$^{0.12}_{0.08}$e-2 &  2.78$^{0.23}_{0.16}$e-26 &  0.592$^{0.013}_{0.013}$ \nl
A\,4059    &   1.19$^{0.09}_{0.06}$e-2 &  2.31$^{0.18}_{0.13}$e-26 &  0.586$^{0.007}_{0.006}$ \nl
0745\,-19    &   6.84$^{0.45}_{0.58}$e-2 &  1.33$^{0.09}_{0.11}$e-25 &  0.563$^{0.004}_{0.004}$ \nl
AWM\,7       &   1.82$^{0.06}_{0.11}$e-2 &  3.54$^{0.12}_{0.21}$e-26 &  0.566$^{0.006}_{0.006}$ \nl
Cygnus\,A       &   1.29$^{0.36}_{0.18}$e-1 &  2.52$^{0.71}_{0.34}$e-25 &  0.614$^{0.021}_{0.021}$ \nl
MKW\,3S      &   1.87$^{0.14}_{0.10}$e-2 &  3.65$^{0.27}_{0.20}$e-26 &  0.579$^{0.015}_{0.015}$ \nl
Ophiuchus       &   1.58$^{0.25}_{0.14}$e-2 &  3.08$^{0.48}_{0.27}$e-26 &  0.546$^{0.009}_{0.009}$ \nl
Tria Aust       &   6.57$^{1.22}_{0.62}$e-3 &  1.28$^{0.24}_{0.12}$e-26 &  0.570$^{0.014}_{0.014}$ \nl
\enddata
\label{ICMresults}
\end{deluxetable}

\begin{deluxetable}{lcccccccc}
\tablewidth{0pt}
\tablecaption{ICM Masses and Mass Fractions}
\tablehead{
							&
\multicolumn{2}{c}{$r_{lim}=1h^{-1}_{50}$~Mpc}	&
\multicolumn{3}{c}{Virial Relation at \rfive}		&
\multicolumn{3}{c}{$\beta$ Model at \rfive}	
\cr
			&
\colhead{\MICM}		&
\colhead{\fICM}		&
\colhead{$r_{lim}$}	&
\colhead{\MICM}		&
\colhead{\fICM}		&
\colhead{$r_{lim}$}	&
\colhead{\MICM}		&	
\colhead{\fICM}
\cr
\colhead{Cluster} 			&
\colhead{$h_{50}^{-5/2}$~M$_\odot$} 	&
\colhead{$h_{50}^{-3/2}$} 		& 
\colhead{$h_{50}^{-1}$~Mpc} 		& 
\colhead{$h_{50}^{-5/2}$~M$_\odot$} 	&
\colhead{$h_{50}^{-3/2}$} 		& 
\colhead{$h_{50}^{-1}$~Mpc} 		& 
\colhead{$h_{50}^{-5/2}$~M$_\odot$} 	&
\colhead{$h_{50}^{-3/2}$}
}
\scriptsize
\startdata
A\,85      &   7.95$^{0.12}_{0.12}$e+13 &  0.194$^{0.006}_{0.007}$ &  1.873 &  1.83$^{0.03}_{0.03}$e+14 &  0.192$^{0.005}_{0.005}$ &  1.719 &  1.64$^{0.04}_{0.03}$e+14 &  0.223$^{0.009}_{0.009}$ \nl
A\,119     &   5.42$^{0.06}_{0.06}$e+13 &  0.158$^{0.010}_{0.010}$ &  1.826 &  1.41$^{0.05}_{0.05}$e+14 &  0.160$^{0.010}_{0.010}$ &  1.635 &  1.20$^{0.05}_{0.05}$e+14 &  0.190$^{0.013}_{0.013}$ \nl
A\,262     &   2.44$^{0.06}_{0.06}$e+13 &  0.168$^{0.011}_{0.010}$ &  1.177 &  3.08$^{0.11}_{0.11}$e+13 &  0.130$^{0.005}_{0.005}$ &  0.998 &  2.42$^{0.05}_{0.06}$e+13 &  0.168$^{0.010}_{0.010}$ \nl
A\,401     &   1.02$^{0.02}_{0.02}$e+14 &  0.194$^{0.008}_{0.008}$ &  2.185 &  3.06$^{0.08}_{0.08}$e+14 &  0.202$^{0.009}_{0.009}$ &  1.934 &  2.60$^{0.07}_{0.07}$e+14 &  0.247$^{0.012}_{0.012}$ \nl
A\,426     &   9.33$^{0.07}_{0.07}$e+13 &  0.207$^{0.022}_{0.021}$ &  1.908 &  2.08$^{0.08}_{0.07}$e+14 &  0.206$^{0.010}_{0.009}$ &  1.849 &  2.01$^{0.04}_{0.04}$e+14 &  0.219$^{0.016}_{0.015}$ \nl
A\,478     &   1.02$^{0.02}_{0.02}$e+14 &  0.204$^{0.006}_{0.006}$ &  1.983 &  2.24$^{0.05}_{0.05}$e+14 &  0.197$^{0.006}_{0.006}$ &  1.901 &  2.14$^{0.04}_{0.05}$e+14 &  0.214$^{0.012}_{0.011}$ \nl
A\,496     &   4.93$^{0.06}_{0.06}$e+13 &  0.184$^{0.006}_{0.006}$ &  1.500 &  8.32$^{0.17}_{0.17}$e+13 &  0.170$^{0.004}_{0.004}$ &  1.369 &  7.42$^{0.12}_{0.11}$e+13 &  0.199$^{0.008}_{0.008}$ \nl
A\,644     &   7.85$^{0.20}_{0.20}$e+13 &  0.169$^{0.006}_{0.006}$ &  1.947 &  1.77$^{0.06}_{0.06}$e+14 &  0.165$^{0.006}_{0.006}$ &  1.805 &  1.62$^{0.03}_{0.03}$e+14 &  0.190$^{0.018}_{0.018}$ \nl
A\,754     &   8.72$^{0.89}_{0.89}$e+13 &  0.172$^{0.007}_{0.007}$ &  2.211 &  2.91$^{0.31}_{0.31}$e+14 &  0.185$^{0.020}_{0.020}$ &  1.953 &  2.45$^{0.44}_{0.44}$e+14 &  0.226$^{0.136}_{0.136}$ \nl
A\,780     &   5.78$^{0.05}_{0.05}$e+13 &  0.186$^{0.008}_{0.008}$ &  1.478 &  8.79$^{0.18}_{0.18}$e+13 &  0.187$^{0.007}_{0.006}$ &  1.464 &  8.70$^{0.17}_{0.18}$e+13 &  0.191$^{0.009}_{0.009}$ \nl
A\,1060    &   1.99$^{0.05}_{0.05}$e+13 &  0.085$^{0.008}_{0.007}$ &  1.335 &  2.76$^{0.15}_{0.15}$e+13 &  0.080$^{0.005}_{0.005}$ &  1.273 &  2.62$^{0.08}_{0.08}$e+13 &  0.087$^{0.009}_{0.009}$ \nl
A\,1367    &   3.71$^{0.06}_{0.06}$e+13 &  0.179$^{0.007}_{0.007}$ &  1.419 &  6.50$^{0.15}_{0.15}$e+13 &  0.157$^{0.006}_{0.006}$ &  1.216 &  5.11$^{0.28}_{0.28}$e+13 &  0.196$^{0.009}_{0.009}$ \nl
A\,1651    &   7.58$^{0.10}_{0.09}$e+13 &  0.182$^{0.010}_{0.010}$ &  1.903 &  1.76$^{0.05}_{0.05}$e+14 &  0.176$^{0.009}_{0.009}$ &  1.706 &  1.53$^{0.04}_{0.04}$e+14 &  0.212$^{0.012}_{0.012}$ \nl
A\,1656    &   8.31$^{0.08}_{0.08}$e+13 &  0.150$^{0.005}_{0.005}$ &  2.173 &  2.39$^{0.11}_{0.11}$e+14 &  0.160$^{0.008}_{0.008}$ &  2.056 &  2.23$^{0.03}_{0.03}$e+14 &  0.177$^{0.019}_{0.019}$ \nl
A\,1689    &   1.18$^{0.02}_{0.03}$e+14 &  0.166$^{0.070}_{0.069}$ &  2.410 &  3.33$^{0.72}_{0.71}$e+14 &  0.164$^{0.069}_{0.069}$ &  2.221 &  3.04$^{0.64}_{0.64}$e+14 &  0.191$^{0.083}_{0.080}$ \nl
A\,1795    &   7.96$^{0.06}_{0.06}$e+13 &  0.197$^{0.005}_{0.005}$ &  1.752 &  1.48$^{0.02}_{0.02}$e+14 &  0.190$^{0.004}_{0.004}$ &  1.752 &  1.48$^{0.03}_{0.03}$e+14 &  0.190$^{0.008}_{0.008}$ \nl
A\,2029    &   1.06$^{0.01}_{0.01}$e+14 &  0.175$^{0.005}_{0.005}$ &  2.237 &  2.78$^{0.06}_{0.06}$e+14 &  0.171$^{0.005}_{0.005}$ &  2.128 &  2.63$^{0.05}_{0.05}$e+14 &  0.188$^{0.010}_{0.010}$ \nl
A\,2052    &   3.63$^{0.09}_{0.10}$e+13 &  0.156$^{0.020}_{0.019}$ &  1.335 &  5.07$^{0.31}_{0.31}$e+13 &  0.146$^{0.013}_{0.012}$ &  1.276 &  4.81$^{0.20}_{0.19}$e+13 &  0.160$^{0.026}_{0.025}$ \nl
A\,2063    &   3.99$^{0.05}_{0.05}$e+13 &  0.127$^{0.021}_{0.020}$ &  1.536 &  6.76$^{0.55}_{0.55}$e+13 &  0.129$^{0.020}_{0.020}$ &  1.458 &  6.36$^{0.50}_{0.50}$e+13 &  0.141$^{0.025}_{0.024}$ \nl
A\,2142    &   1.29$^{0.06}_{0.06}$e+14 &  0.221$^{0.007}_{0.006}$ &  2.234 &  3.70$^{0.04}_{0.04}$e+14 &  0.229$^{0.004}_{0.004}$ &  2.206 &  3.65$^{0.22}_{0.24}$e+14 &  0.227$^{0.024}_{0.017}$ \nl
A\,2199    &   4.77$^{0.03}_{0.03}$e+13 &  0.151$^{0.010}_{0.009}$ &  1.609 &  8.40$^{0.16}_{0.16}$e+13 &  0.139$^{0.005}_{0.005}$ &  1.493 &  7.71$^{0.14}_{0.14}$e+13 &  0.159$^{0.006}_{0.006}$ \nl
A\,2204    &   1.17$^{0.04}_{0.04}$e+14 &  0.211$^{0.010}_{0.011}$ &  2.275 &  3.52$^{0.22}_{0.23}$e+14 &  0.206$^{0.013}_{0.014}$ &  1.979 &  2.93$^{0.12}_{0.14}$e+14 &  0.260$^{0.028}_{0.032}$ \nl
A\,2244    &   7.49$^{0.26}_{0.25}$e+13 &  0.164$^{0.048}_{0.047}$ &  2.021 &  1.89$^{0.29}_{0.29}$e+14 &  0.158$^{0.046}_{0.045}$ &  1.782 &  1.61$^{0.23}_{0.23}$e+14 &  0.196$^{0.061}_{0.060}$ \nl
A\,2255    &   7.13$^{0.09}_{0.09}$e+13 &  0.150$^{0.030}_{0.030}$ &  2.049 &  2.02$^{0.20}_{0.20}$e+14 &  0.162$^{0.032}_{0.032}$ &  2.008 &  1.97$^{0.20}_{0.20}$e+14 &  0.168$^{0.035}_{0.034}$ \nl
A\,2256    &   9.17$^{0.16}_{0.16}$e+13 &  0.166$^{0.003}_{0.003}$ &  2.078 &  2.32$^{0.06}_{0.06}$e+14 &  0.178$^{0.005}_{0.005}$ &  2.113 &  2.36$^{0.03}_{0.03}$e+14 &  0.172$^{0.016}_{0.016}$ \nl
A\,2319    &   1.10$^{0.05}_{0.05}$e+14 &  0.213$^{0.004}_{0.004}$ &  2.290 &  3.95$^{0.12}_{0.12}$e+14 &  0.226$^{0.007}_{0.007}$ &  1.911 &  3.01$^{0.20}_{0.20}$e+14 &  0.297$^{0.027}_{0.027}$ \nl
A\,2597    &   4.99$^{0.17}_{0.17}$e+13 &  0.206$^{0.015}_{0.015}$ &  1.439 &  7.73$^{0.48}_{0.48}$e+13 &  0.178$^{0.012}_{0.012}$ &  1.290 &  6.78$^{0.27}_{0.27}$e+13 &  0.217$^{0.020}_{0.020}$ \nl
A\,3112    &   5.84$^{0.16}_{0.16}$e+13 &  0.202$^{0.015}_{0.015}$ &  1.644 &  1.14$^{0.07}_{0.07}$e+14 &  0.177$^{0.014}_{0.014}$ &  1.412 &  9.30$^{0.33}_{0.33}$e+13 &  0.227$^{0.027}_{0.027}$ \nl
A\,3158    &   6.89$^{0.12}_{0.12}$e+13 &  0.185$^{0.014}_{0.014}$ &  1.778 &  1.49$^{0.06}_{0.06}$e+14 &  0.183$^{0.013}_{0.013}$ &  1.632 &  1.34$^{0.06}_{0.06}$e+14 &  0.212$^{0.020}_{0.019}$ \nl
A\,3266    &   8.40$^{0.15}_{0.15}$e+13 &  0.206$^{0.016}_{0.016}$ &  1.888 &  2.11$^{0.09}_{0.09}$e+14 &  0.216$^{0.017}_{0.017}$ &  1.801 &  1.99$^{0.09}_{0.09}$e+14 &  0.234$^{0.021}_{0.021}$ \nl
A\,3391    &   4.83$^{0.12}_{0.12}$e+13 &  0.149$^{0.014}_{0.013}$ &  1.810 &  1.22$^{0.07}_{0.07}$e+14 &  0.141$^{0.012}_{0.012}$ &  1.511 &  9.24$^{0.59}_{0.56}$e+13 &  0.184$^{0.021}_{0.020}$ \nl
A\,3526    &   2.66$^{0.06}_{0.06}$e+13 &  0.122$^{0.010}_{0.009}$ &  1.427 &  4.38$^{0.21}_{0.20}$e+13 &  0.104$^{0.005}_{0.005}$ &  1.227 &  3.55$^{0.07}_{0.06}$e+13 &  0.132$^{0.012}_{0.012}$ \nl
A\,3532    &   4.57$^{0.13}_{0.13}$e+13 &  0.170$^{0.013}_{0.015}$ &  1.591 &  9.13$^{0.36}_{0.39}$e+13 &  0.156$^{0.006}_{0.007}$ &  1.378 &  7.42$^{0.59}_{0.53}$e+13 &  0.195$^{0.018}_{0.022}$ \nl
A\,3558    &   7.37$^{0.16}_{0.16}$e+13 &  0.223$^{0.006}_{0.006}$ &  1.810 &  1.81$^{0.03}_{0.03}$e+14 &  0.210$^{0.005}_{0.005}$ &  1.524 &  1.41$^{0.05}_{0.05}$e+14 &  0.273$^{0.013}_{0.013}$ \nl
A\,3562    &   4.75$^{0.11}_{0.11}$e+13 &  0.244$^{0.034}_{0.034}$ &  1.478 &  9.04$^{0.65}_{0.65}$e+13 &  0.193$^{0.026}_{0.026}$ &  1.157 &  6.05$^{0.44}_{0.44}$e+13 &  0.269$^{0.037}_{0.037}$ \nl
A\,3571    &   7.93$^{0.09}_{0.09}$e+13 &  0.176$^{0.006}_{0.006}$ &  1.992 &  1.99$^{0.06}_{0.06}$e+14 &  0.173$^{0.007}_{0.007}$ &  1.776 &  1.72$^{0.03}_{0.03}$e+14 &  0.211$^{0.013}_{0.013}$ \nl
A\,3667    &   8.16$^{0.10}_{0.10}$e+13 &  0.209$^{0.011}_{0.011}$ &  2.006 &  2.46$^{0.07}_{0.07}$e+14 &  0.209$^{0.011}_{0.011}$ &  1.673 &  1.86$^{0.07}_{0.07}$e+14 &  0.274$^{0.015}_{0.015}$ \nl
A\,4038    &   3.48$^{0.08}_{0.08}$e+13 &  0.179$^{0.014}_{0.014}$ &  1.378 &  5.48$^{0.36}_{0.36}$e+13 &  0.144$^{0.009}_{0.009}$ &  1.156 &  4.27$^{0.07}_{0.07}$e+13 &  0.190$^{0.020}_{0.020}$ \nl
A\,4059    &   4.24$^{0.07}_{0.07}$e+13 &  0.170$^{0.010}_{0.010}$ &  1.536 &  7.68$^{0.28}_{0.28}$e+13 &  0.146$^{0.008}_{0.008}$ &  1.313 &  6.19$^{0.17}_{0.17}$e+13 &  0.188$^{0.013}_{0.012}$ \nl
0745\,-19    &   1.20$^{0.03}_{0.03}$e+14 &  0.220$^{0.027}_{0.027}$ &  2.211 &  3.31$^{0.23}_{0.22}$e+14 &  0.211$^{0.026}_{0.026}$ &  1.940 &  2.81$^{0.18}_{0.18}$e+14 &  0.264$^{0.034}_{0.033}$ \nl
AWM\,7       &   3.84$^{0.04}_{0.04}$e+13 &  0.137$^{0.007}_{0.008}$ &  1.498 &  6.29$^{0.23}_{0.23}$e+13 &  0.129$^{0.006}_{0.006}$ &  1.400 &  5.81$^{0.13}_{0.13}$e+13 &  0.146$^{0.010}_{0.010}$ \nl
Cygnus\,A       &   7.70$^{0.76}_{0.75}$e+13 &  0.229$^{0.015}_{0.015}$ &  1.933 &  2.19$^{0.11}_{0.11}$e+14 &  0.209$^{0.015}_{0.015}$ &  1.523 &  1.50$^{0.14}_{0.14}$e+14 &  0.292$^{0.032}_{0.032}$ \nl
MKW\,3S      &   3.81$^{0.10}_{0.10}$e+13 &  0.176$^{0.017}_{0.016}$ &  1.419 &  6.13$^{0.47}_{0.47}$e+13 &  0.148$^{0.012}_{0.012}$ &  1.219 &  4.98$^{0.15}_{0.15}$e+13 &  0.189$^{0.025}_{0.024}$ \nl
Ophiuchus       &   9.85$^{0.12}_{0.12}$e+13 &  0.134$^{0.010}_{0.010}$ &  2.374 &  2.75$^{0.15}_{0.15}$e+14 &  0.142$^{0.011}_{0.011}$ &  2.270 &  2.62$^{0.10}_{0.10}$e+14 &  0.154$^{0.016}_{0.016}$ \nl
Tria Aust       &   1.08$^{0.02}_{0.02}$e+14 &  0.173$^{0.015}_{0.013}$ &  2.337 &  3.37$^{0.18}_{0.18}$e+14 &  0.182$^{0.012}_{0.012}$ &  2.340 &  3.38$^{0.11}_{0.12}$e+14 &  0.181$^{0.024}_{0.022}$ \nl
\enddata
\label{ICMmasses}
\end{deluxetable}

\end{document}